\documentclass[3p]{elsarticle}

\usepackage[utf8]{inputenc}
\usepackage{amsmath,amssymb,amsfonts}
\usepackage{graphicx}
\usepackage{booktabs}
\usepackage{tabularx}
\usepackage{hyperref}
\usepackage{url}
\usepackage{xcolor}
\usepackage{enumitem}
\usepackage{listings}          

\newcommand{\xmark}{$\times$}

\newcommand{\jaxdae}{jaxdae}

\begin{document}

\begin{frontmatter}

\title{jaxdae: A JAX-native Differentiable Solver for Differential-Algebraic Equations in Coupled Multi-physics}

\author[a,b]{Chengyuan Li}

\author[b]{Shanfang Huang\corref{cor1}}
\ead{sfhuang@mail.tsinghua.edu.cn}

\author[a]{Jian Deng}

\cortext[cor1]{Corresponding author}

\address[a]{National Key Laboratory of Nuclear Reactor Technology, Nuclear Power Institute of China, Chengdu, 610213, China}
\address[b]{Department of Engineering Physics, Tsinghua University, Beijing, 100084, China}

\begin{abstract}
Many engineered models begin as partial differential equations. Spatial discretization converts them into ordinary differential equations coupled to algebraic constraints---conservation closures, constitutive laws, network topology---whose joint evolution is a differential-algebraic equation (DAE). Parameter inversion, uncertainty quantification, Bayesian inference, and optimal control all require gradients of this solve. The two software traditions that should supply them have not met: industrial acausal modeling tools simulate DAEs forward but stop at reverse-mode differentiation, while differentiable-physics frameworks in JAX handle explicit ODEs and PDEs but leave the algebraic-constraint layer untouched. Here we show that the forward DAE solve and its reverse-mode sensitivity can be unified in one JAX-native suite. jaxdae pairs adaptive BDF, Radau, and Rosenbrock integration with Pantelides index reduction and dummy derivatives, and makes the adaptive BDF path differentiable by freezing the accepted step grid and re-solving a variable-step BDF-2 on it for the backward pass. The full pipeline differentiates under one \texttt{jax.grad} call, XLA fuses a batched parameter sweep into one program whose wall time stays nearly flat from batch~1 to~1000, and the DAE becomes a differentiable primitive for inference, control, and design.
\end{abstract}


\end{frontmatter}

\vspace{0.5em}
\noindent\textbf{\large Program Summary}

\vspace{0.5em}
\noindent\textit{Program Title:} jaxdae v0.28.0 --- JAX-native differentiable DAE solver

\noindent\textit{CPC Library link to program files:} The code archive will be
deposited in the CPC Program Library (hosted on Mendeley Data) and assigned
a catalog number upon acceptance.

\noindent\textit{Licensing provisions:} Apache-2.0

\noindent\textit{Programming language:} Python ($\geq$3.11), JAX ($\geq$0.9, $<$0.11)

\noindent\textit{Supplementary material:} None. The full source code, benchmark
suites, pre-computed data, and a reproduction manifest (\texttt{paper\_manifest.md})
are deposited alongside this manuscript in the CPC Program Library.

\noindent\textit{Journal Reference of previous version:} None.

\noindent\textit{Does the new version supersede the previous version?} Not
applicable; this is the first release.

\noindent\textit{Reasons for the new version:} Not applicable.

\noindent\textit{Summary of revisions:} Not applicable.

\vspace{0.5em}
\noindent\textit{Nature of problem:}
Many engineered systems couple time-evolving states with algebraic constraints that must hold instantaneously. These take the semi-explicit mass-matrix form $M(t,y)\,\dot{y}=f(t,y,p)$ together with $0=g(t,y,p)$, where $y$ collects the states and $p$ the parameters. Some equations are implicit in $\dot{y}$, and the differential index can exceed one, so naive time stepping drifts off the constraint manifold and eventually diverges. The task is to advance $y(t)$ while keeping every constraint satisfied to integration tolerance, for assembled plants that couple hundreds of variables and span milliseconds to hours.

\vspace{0.5em}
\noindent\textit{Solution method:}
jaxdae advances the residual with adaptive implicit integration---BDF(1--5), Radau IIA, and Rosenbrock--Wanner---driven by a PI step-size and order controller. High-index input is reduced before integration: Pantelides' structural analysis marks which equations to differentiate, and dummy derivatives restore a nonsingular augmented system. For differentiation, a custom reverse-mode rule freezes the accepted step grid and re-solves a variable-step BDF-2 on it with an implicit-function-theorem adjoint, so the adaptive controller never enters cotangent propagation. The differentiable paths are the BDF scan/replay-adjoint path and the fixed-step ROS2 path; Radau5 is provided as a forward-only solver. Every primitive is pure JAX, so \texttt{jax.grad}, \texttt{vmap}, and \texttt{jit} compose across these differentiable paths.

\vspace{0.5em}
\noindent\textit{Additional comments including restrictions and unusual features:}
jaxdae assumes the DAE is index-1 after reduction: Pantelides and dummy derivatives accept higher-index input, but the integrator operates on the reduced system. Typed connectors spanning electrical, thermal, fluid, and mechanical domains assemble through \texttt{connect} semantics in the Modelica style. Restrictions: no distributed multi-GPU execution, and a subset of state-event types is verified numerically rather than by analytic jump-condition matching. The solver integrates with Optax, NumPyro, and Flax, and a test suite of 3{,}205 cases collected by \texttt{pytest -m "" -{}-timeout=600} (3{,}177 passing at the pinned commit; the remaining 28 --- 12 known failures reproducing identically at v0.27.1, 9 skipped, 7 xfailed --- are documented in \texttt{RELEASE\_NOTES.md}) covers forward integration, index reduction, gradient correctness, and event handling. As a robustness feature, \texttt{solve\_dae} checks the initial residual at call time and repairs inconsistent initial conditions automatically (with a \texttt{RuntimeWarning}) rather than failing on them.

\vspace{0.5em}
\noindent\textit{References:}
Brenan, Campbell, and Petzold~\cite{brenan1996dae}; Ascher and Petzold~\cite{ascher1998dae}; Pantelides~\cite{pantelides1988}; Mattsson and Söderlind~\cite{mattssonsoderlind1993}; Cao, Li, Petzold, and Serban~\cite{caolipetzold2003}.

\vspace{1em}\hrule\vspace{1em}

\section{Introduction}
\label{sec:intro}

Beneath every simulation of a physical system lies a partial differential equation. Mass, momentum, energy, charge, and species each obey a balance law that pairs storage in a control volume with flux across its boundary~\cite{leveque2002fvm}. A digital computer represents each field by sampling it at finitely many points, so the first step of any simulation is spatial discretization: finite differences, finite volumes, finite elements, or lumped-parameter networks replace the field with a finite set of nodal values. After this step the differential variables---node temperatures, pressures, compositions---satisfy ordinary differential equations in time. Discretization does not, however, eliminate every relation that the continuum imposed. Conservation at junctions, constitutive laws, equation-of-state closures, network topology, and rigid-link conditions survive as constraints that must hold at every instant alongside the time evolution~\cite{brenan1996dae,ascher1998dae}. The object that results---ordinary differential equations coupled to algebraic constraints---is a differential-algebraic equation (DAE), and it is the form in which most coupled multi-physics is actually integrated.
Forward simulation is only the first ask. The questions that drive engineering all reduce to optimization over the same physical model: what parameter values fit the data, which design tolerances matter, how a controller should respond, what an experiment reveals~\cite{tarantola2005inverse}. Parameter inversion matches a model to measurements; uncertainty quantification propagates input uncertainties through the solve; Bayesian inference updates a posterior over parameters from a likelihood defined by the residual; optimal control and model-predictive control search over actuator trajectories; sensitivity analysis ranks which inputs a safety margin depends on. Each task demands the same primitive: the gradient of the solver output with respect to its input parameters. Automatic differentiation~\cite{baydin2018ad,margossian2019adreview} supplies this gradient at machine precision and turns a forward model into an inference engine.
The gradient primitive is where the tooling fractures. Differentiable simulators have recast molecular dynamics~\cite{schoenholz2020jaxmd}, soft- and rigid-body animation~\cite{hu2020difftaichi,bezgin2023jaxfluids}, finite-element mechanics~\cite{xue2023jaxfem}, and lattice Boltzmann flow~\cite{ataei2024xlb} as programs whose derivatives come from reverse-mode automatic differentiation rather than hand-derived adjoints~\cite{giles2002adjoint}. These efforts, however, target systems written as explicit ODEs or PDEs and leave the algebraic-constraint layer of a DAE untouched. The industrial tools that do solve DAEs took a different route decades earlier and have not adopted end-to-end differentiation.
That industrial route is the acausal modeling tradition. Languages formalized in the Modelica standard~\cite{modelica_spec} and popularized through textbooks and tools~\cite{tiller2001modelica} let engineers assemble a plant from typed components joined by effort--flow pairs and compile the result into a DAE residual. A turbine, a heat exchanger, and a controller become blocks linked by \texttt{connect} equations; the compiler flattens them into one mass-matrix system. This compositional style is why Modelica ecosystems dominate energy, automotive, and aerospace design. The compiled DAE, though, is built to be simulated forward, not differentiated end to end.
The two currents need a common substrate, and JAX~\cite{bradbury2018jax} supplies one. Its functional core---pure functions traced to XLA---turns reverse-mode AD, vectorization (\texttt{vmap}), and just-in-time compilation into composable transformations of the same program. On this substrate, Neural ODEs reframed time evolution as a differentiable layer~\cite{chen2018neuralode,kidger2022nodf}, diffrax~\cite{kidger2021diffrax} gathered a library of adaptive solvers for ODEs, stochastic differential equations (SDEs), and controlled differential equations (CDEs), and supporting libraries for linear algebra~\cite{rader2023lineax} and root-finding~\cite{rader2024optimistix} made the implicit solves that stiff systems require composable and differentiable. What JAX still lacks is a solver for the equation class---DAEs---that industry actually writes.
Neural-DAE research has meanwhile proliferated on top of ad hoc, hand-built pipelines. Model-integrated neural networks embed a battery electrochemical DAE in a recurrent cell and learn the algebraic root directly~\cite{huang2024minn}. Kolmogorov--Arnold networks solve index-3 DAEs by collocation on the residual~\cite{luo2025daekan}. A multiscale Lagrangian formulation separates fast and slow dynamics to learn stiff multibody systems~\cite{huang2024multiscale}, and an operator-splitting scheme learns unknown component laws while enforcing known algebraic constraints~\cite{koch2024neuraldae}. These works confirm the demand: each wants gradients through a constrained dynamical system, and each builds its own projection layer, integrator, or penalty loss to obtain them. They differ in philosophy---soft penalty versus hard projection, a priori fitting versus a posteriori rollout---but share one unmet need, a reusable, differentiable, index-reducing DAE solver. jaxdae targets exactly that layer.
DAEs differ from ODEs by more than the presence of constraints. The differential index counts how many times the algebraic equations must be differentiated before the system reduces to an explicit ODE~\cite{brenan1996dae}. Index-1 systems are benign: the constraint Jacobian is invertible and the algebraic variables follow directly. Higher-index systems are not. Each differentiation of a constraint amplifies numerical error, so a naive integrator drifts off the constraint manifold and eventually diverges. Stiffness compounds the difficulty: coupled multi-physics routinely mixes fast transients with slow drift, separated by many orders of magnitude in time scale, a phenomenon first identified by Curtiss and Hirschfelder~\cite{curtiss1952stiff}, and only implicit methods stay stable~\cite{hairerwanner1996ii}.
Index reduction is therefore mandatory before integration. Pantelides' structural algorithm decides, by bipartite matching on the equation sparsity pattern, which equations to differentiate and how to construct consistent initial conditions~\cite{pantelides1988}. Pryce's signature-matrix method generalizes this to a broader class of systems~\cite{pryce2001structural,nedialkov2015daesa}. Dummy derivatives then restore a nonsingular augmented system without altering the solution trajectory~\cite{mattssonsoderlind1993}. For mechanical systems the Gear--Gupta--Leimkuhler stabilization~\cite{gear1985ggl} reduces index~3 to index~2 by enforcing position, velocity, and acceleration constraints simultaneously. These structural and stabilization steps are the entry ticket to robust high-index simulation, yet they are delicate and graph-theoretic, and they are rarely expressed in a differentiable form---which is why most differentiable-physics frameworks sidestep high-index DAEs altogether.
The solvers that industry trusts do handle DAEs. SUNDIALS IDA and its predecessor DASSL integrate stiff, high-index systems in Fortran and C~\cite{hindmarsh2005sundials}. They are fast, validated, and battle-tested in reactor-safety and process-simulation codes. Their cost is an integration model that predates AD: a user writes residual callbacks in C, links against a fixed library, and accepts a black-box forward solve.
Obtaining a gradient from such a solver means deriving an adjoint by hand. The DAE adjoint sensitivity equations are well understood~\cite{caolipetzold2003,petzold2006sensitivity,tolsma1997sensitivity}, and production adjoint solvers such as PETSc TSAdjoint~\cite{zhang2022tsadjoint} implement discrete adjoints for time-dependent problems with checkpointing~\cite{griewank2000revolve}. The theoretical subtleties of discrete versus continuous adjoints are well studied~\cite{giles2002adjoint,sanzserna2016symplectic}. But encoding an adjoint for every new DAE model, managing checkpointing, and threading it through a controller that adapts step size is months of expert work. Modern implicit-differentiation frameworks~\cite{blondel2021implicit,margossian2021implicit} automate the IFT gradient of a fixed-point solve but do not address the adaptive time-stepping and index-reduction pipeline of a DAE integrator. The practical outcome is a hard split: simulate forward in a trusted classical solver without gradients, or differentiate an ODE surrogate and lose the constraint structure. No widely used path does both at once.
Recent work has begun to close this split, and Table~\ref{tab:competitors} maps the landscape. diffrax~\cite{kidger2021diffrax} remains ODE-only; its DAE request has stayed open since 2022~\cite{diffrax_issue62}. A concurrent PyTorch counterpart, torchdae~\cite{torchdae2026}, integrates DAEs but ships hand-written adjoint code that neither vectorizes nor composes with \texttt{torch.vmap}. ModelingToolkit.jl~\cite{ma2021modelingtoolkit}, in the Julia ecosystem, performs acausal assembly, index reduction, and reverse-mode AD through the SciML sensitivity stack~\cite{rackauckas2021ude,ma2021casavsaad}, yet its batched path relies on ensemble threading rather than XLA fusion. Classical options complete the picture: SUNDIALS IDA~\cite{hindmarsh2005sundials} solves but differentiates only through a manual adjoint, and CasADi~\cite{andersson2019casadi} offers symbolic sensitivities restricted to index-1. PyBaMM~\cite{sulzer2021pybamm} ships a JAX-native differentiable DAE solver for battery models, but it is domain-specific and does not perform index reduction or acausal assembly. Under the versions inspected (Table~\ref{tab:competitors}), no existing tool combines a differentiable DAE solve, automatic index reduction, and the fused, vectorized batched gradient that JAX compiles.

\begin{table}[htbp]
\centering
\caption{Capability comparison of differentiable and classical DAE solvers.
Capabilities were verified by source-code inspection at the versions
listed below (accessed July 2026). ``Reverse AD'' marks whether a
reverse-mode gradient through the solve is available without hand-written
adjoint code.}
\label{tab:competitors}
\begin{tabular}{l l c c c c}
\toprule
Tool & Ecosystem & DAE & Index red. & Reverse AD & vmap batch \\
\midrule
\textbf{jaxdae} & JAX & \checkmark & \checkmark & \checkmark (one-line) & \checkmark (fused) \\
diffrax & JAX & \xmark (\#62) & \xmark & ODE only & N/A \\
torchdae & PyTorch & \checkmark & \checkmark & hand-written & fwd\checkmark/grad\xmark \\
ModelingToolkit.jl & Julia & \checkmark & \checkmark (+tearing) & \checkmark (Zygote) & ensemble (15$\times$) \\
SUNDIALS IDA & C & \checkmark & \xmark & manual ($\sim$95 lines) & \xmark \\
CasADi & Python & \checkmark (idx-1) & \xmark & symbolic & CPU threads \\
PyBaMM & JAX & \checkmark (battery) & \xmark & \checkmark (battery) & \xmark \\
\bottomrule
\end{tabular}
\\[3pt]
\footnotesize Versions inspected (July 2026): jaxdae v0.28.0 (Program Summary); diffrax (GitHub
\texttt{patrick-kidger/diffrax}, issue \#62); torchdae
(\texttt{yousef-rafat/torchdae}, alpha); ModelingToolkit.jl
(\texttt{SciML/ModelingToolkit.jl}); SUNDIALS IDA 7.5.0; CasADi 3.6
(\texttt{casadi/casadi}). ``Index red.'' includes Pantelides plus
dummy-derivative or structural tearing. ``manual ($\sim$95 lines)'' denotes
the user-written IDA adjoint callback from the SUNDIALS examples.
\end{table}

The broader landscape of differentiable physics in JAX has grown
rapidly. Beyond the ODE solvers cited above, rigid-body engines for
large-scale simulation~\cite{freeman2021brax}, smoothed-particle
hydrodynamics~\cite{toshev2024jaxsph}, and the second-generation
JAX-Fluids framework~\cite{bezgin2024jaxfluids2} extend differentiable
simulation to new equation classes. Outside JAX, accelerator modeling in
Julia~\cite{jutrack2025} and deep-potential training via AD~\cite{chemtrain2025}
show that the demand for differentiable solvers spans languages and
domains; both were published in CPC, underscoring the venue's interest in
this class of software. Neural controlled differential equations extend
the adjoint approach to irregular time series~\cite{kidger2020cde}, but
still target ODE-class dynamics rather than algebraically constrained
systems.
These observations fix four design goals for jaxdae. First, the solver must be native JAX, so that reverse-mode AD, \texttt{vmap}, and \texttt{jit} compose without wrappers. Second, index reduction---Pantelides plus dummy derivatives---must run automatically and stay in the same traced graph as the solve. Third, acausal assembly in the Modelica style must be differentiable end to end, from \texttt{connect} equations to the final loss. Fourth, the result must plug into the surrounding ML stack---Optax optimizers, NumPyro inference, Flax neural closures---as an ordinary differentiable function.
jaxdae realizes these goals in three contributions.
\textbf{C1 (infrastructure).} To the best of our comparison under the versions listed in Table~\ref{tab:competitors}, jaxdae is the first JAX-native DAE solver suite to combine adaptive BDF, Radau, and Rosenbrock integration with Pantelides and dummy-derivative index reduction and Modelica-style acausal assembly of multi-physics connector types. The BDF and fixed-step Rosenbrock paths are differentiable end to end through a single \texttt{jax.grad} call; Radau is provided as a forward-only solver.
\textbf{C2 (realization).} A frozen-grid replay adjoint makes the adaptive BDF path differentiable: the forward solve runs adaptively, and the backward pass re-solves a variable-step BDF-2 on the accepted grid. The decoupling itself follows the controller-decoupling framework of Alexe and Sandu~\cite{alexesandu2009}; our contribution is the first correct, end-to-end \texttt{jax.grad}-compatible JAX realization of this idea for DAEs, which raw adaptive-controller AD cannot provide (its cotangents are silently wrong because the accept/reject branch breaks the chain rule).
\textbf{C3 (ecosystem).} jaxdae fills the structural gap that has kept DAE modeling out of the JAX scientific stack, and its \texttt{vmap}-fused batched gradient makes parameter sweeps and ensemble inference practical on a single GPU.
The remainder of the paper proceeds as follows. Section~\ref{sec:model} states the DAE formulation and defines index and consistency. Section~\ref{sec:numerical} develops the numerical methods: implicit integration, Newton--Krylov iteration, adaptive control, acausal assembly, index reduction, event handling, and the frozen-grid replay adjoint. Section~\ref{sec:software} describes the software design. Section~\ref{sec:validation} validates jaxdae as a classical solver against standard test problems and cross-library references. Section~\ref{sec:performance} reports batched-gradient throughput. Sections~\ref{sec:ad} and~\ref{sec:applications} examine differentiability and three engineering applications, and Section~\ref{sec:conclusion} concludes.

\section{Mathematical Model}
\label{sec:model}

\subsection{DAE formulation}
A differential-algebraic equation (DAE) couples evolution laws with
algebraic constraints that hold at every instant. In its most general
implicit form the model reads
\begin{equation}
F\!\left(t,\, y(t),\, \dot y(t),\, p\right) = 0,
\qquad t\in[t_0,t_f],
\label{eq:dae-general}
\end{equation}
where $y(t)\in\mathbb{R}^{n_d}$ is the state, $p\in\mathbb{R}^{n_p}$
collects parameters, and the residual $F$ may depend on $t$, $y$, and
$\dot y$ without being solvable for $\dot y$ explicitly. When
$\partial F/\partial\dot y$ is singular the system is genuinely
algebraically constrained and cannot be recast as an ODE; this is the
property that separates DAEs from stiff ODEs~\cite{brenan1996dae,ascher1998dae}.
jaxdae targets the semi-explicit, mass-matrix partition that dominates
engineering practice. Splitting the state into a differential component
$y$ and an algebraic component $z$, the model becomes
\begin{align}
M(t,y)\,\dot y &= f(t,y,z,p), \label{eq:mass}\\
0 &= g(t,y,z,p), \label{eq:algebraic},
\end{align}
with $M(t,y)\in\mathbb{R}^{n_d\times n_d}$ the mass matrix on the
differential block (independent of the algebraic block $z$ in the systems
treated here), $f$ the dynamic residual, and $g$ the algebraic
constraint. The mass matrix acts only on the differential variables; the
algebraic rows carry no $\dot z$ term, so $M$ is a block-diagonal
projection rather than a general singular matrix. This is the input
contract of every integrator in this paper: a residual $F(t,x,\dot x,p)$
affine in $\dot x$ together with the boolean differential/algebraic mask;
fully implicit forms outside this contract are rejected at assembly time. The general implicit
form~\eqref{eq:dae-general} reduces to~\eqref{eq:mass}--\eqref{eq:algebraic}
when the residual is affine in $\dot y$ (i.e., $F_{\dot y}$ is
independent of $\dot y$), which covers the standard engineering DAEs.
Equation~\eqref{eq:mass} reduces to a pure ODE when $M$ is nonsingular
and no algebraic constraint is present,
since $\dot y = M^{-1}f$ then eliminates the derivative. The partition
in~\eqref{eq:mass}--\eqref{eq:algebraic} is not merely notational: it
fixes the sparsity pattern that the integrator, the index-reduction
pass, and the adjoint each exploit~\cite{hairerwanner1996ii}.
The parameters $p$ are first-class quantities. They enter $M$, $f$, and
$g$ and are exposed to \texttt{jax.grad} without source transformation,
which is the property that makes the forward solve differentiable
end-to-end. Table~\ref{tab:symbols} fixes the notation used throughout
the paper.
\subsection{Differential index and index reduction}
The differential index measures how far a DAE sits from an explicit ODE.
Formally, the index $\nu$ of~\eqref{eq:dae-general} is the smallest
non-negative integer for which the augmented system obtained by
differentiating the constraint $\nu$ times,
\begin{equation}
F=0,\quad \tfrac{dF}{dt}=0,\quad \ldots,\quad \tfrac{d^{\nu}F}{dt^{\nu}}=0,
\label{eq:index-def}
\end{equation}
determines $\dot y$ as a continuous function of $(t,y)$ via the
implicit-function theorem~\cite{brenan1996dae}. An ODE has index~$0$.
For the general implicit form, index~$1$ holds exactly when
$\partial F/\partial\dot y$ is nonsingular; for the semi-explicit
system~\eqref{eq:mass}--\eqref{eq:algebraic}, index~$1$ reduces to the
nonsingularity of the constraint Jacobian $\partial g/\partial z$, which
lets~\eqref{eq:algebraic} be solved for $z$ given $y$.
Index matters because it controls numerical difficulty. Algebraic
variables of an index-$\nu$ system carry $\nu$ units of hidden
differentiation; a discretization that treats them as differential
states incurs error that grows as $h^{-\nu}$, so the solution drifts off
the constraint manifold and the iteration may fail
outright~\cite{ascher1998dae}. Higher-index DAEs therefore require
index reduction before time integration. Two ingredients suffice: a
structural algorithm that identifies which constraints to
differentiate~\cite{pantelides1988}, and a regularization that removes
the resulting redundancy without changing the solution
set~\cite{mattssonsoderlind1993}. Section~3.5 implements both.
\begin{table}[htbp]
\centering
\caption{Notation. The differential/algebraic split follows the
semi-explicit form~\eqref{eq:mass}--\eqref{eq:algebraic}.}
\label{tab:symbols}
\small
\begin{tabular}{l l}
\toprule
Symbol & Meaning \\
\midrule
$t$            & independent variable (time), $t\in[t_0,t_f]$ \\
$y(t)$         & differential state vector, $y\in\mathbb{R}^{n_d}$ \\
$z(t)$         & algebraic state vector, $z\in\mathbb{R}^{n_a}$ \\
$\dot y$       & time derivative of the differential state \\
$F$            & general DAE residual, Eq.~\eqref{eq:dae-general} \\
$M(t,y)$       & mass matrix on the differential block \\
$f$            & dynamic residual, Eq.~\eqref{eq:mass} \\
$g$            & algebraic constraint, Eq.~\eqref{eq:algebraic} \\
$p$            & parameter vector exposed to \texttt{jax.grad} \\
$h_k$          & accepted step size at step $k$ \\
$\lambda$      & adjoint costate, Sections~3.6--3.7 \\
$J$            & iteration Jacobian $\partial R/\partial x$ \\
$\sigma_k$     & scaled local error estimate at step $k$ \\
\bottomrule
\end{tabular}
\end{table}
\section{Numerical Methods}
\label{sec:numerical}

\begin{figure}[htbp]
\centering
\includegraphics[width=0.95\textwidth]{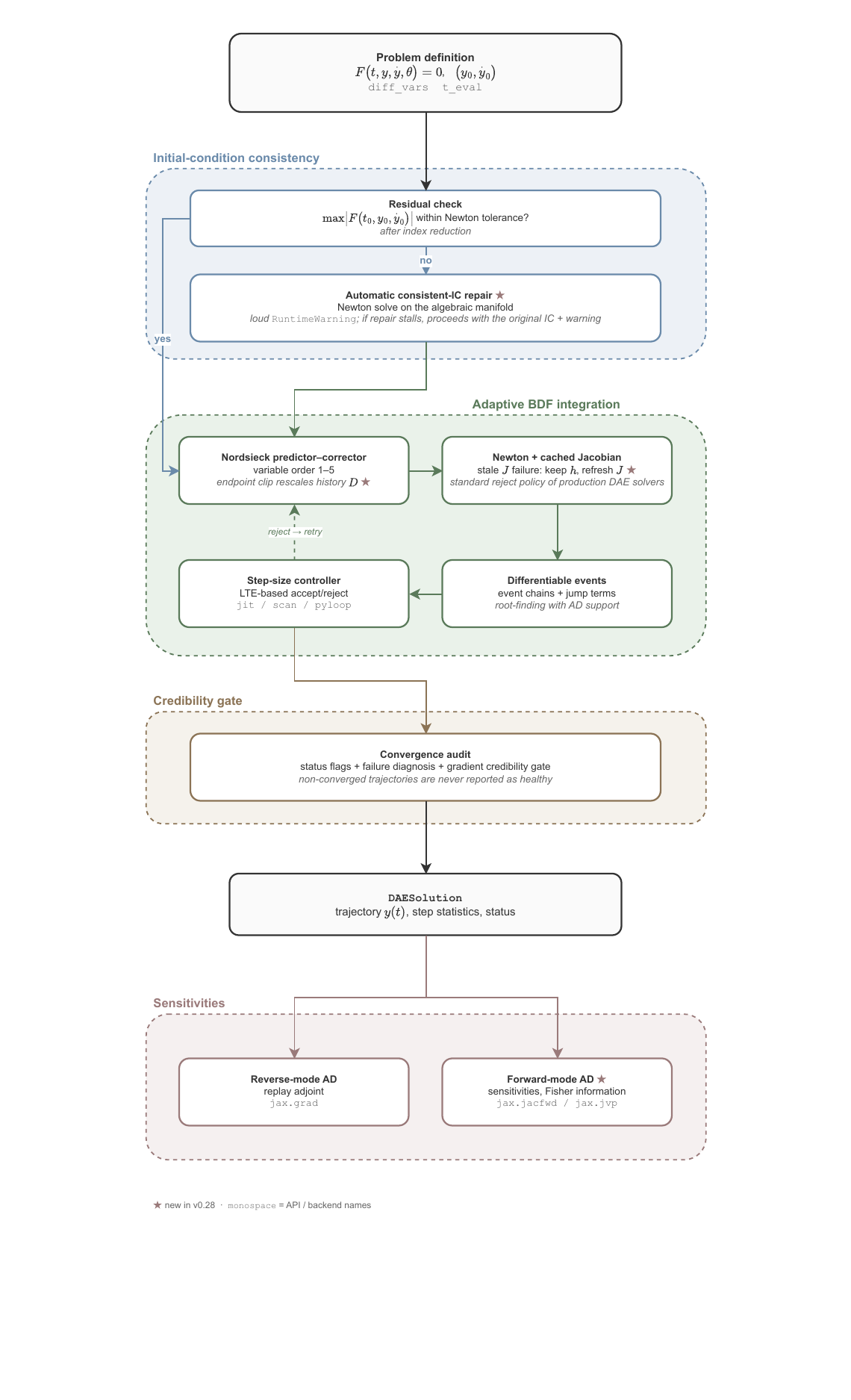}
\caption{Main solve pipeline of jaxdae. A DAE problem is defined by
acausal assembly or by a direct residual, checked and repaired by the
initial-condition consistency layer, integrated by the adaptive BDF
controller, and passed through the credibility gates (order
verification, cross-library comparison) before producing trajectory
output and forward- or reverse-mode sensitivities.}
\label{fig:main_flow}
\end{figure}

Figure~\ref{fig:main_flow} gives an overview of the solve pipeline that
this section develops piece by piece: a problem is defined through the
acausal assembly of Section~3.4, passes an initial-condition consistency
layer (Section~4.2), is integrated by the adaptive implicit methods of
Sections~3.1--3.3 under the credibility gates of Section~5, and yields
both the trajectory output and the differentiable sensitivities of
Section~3.7.

\subsection{Implicit time integration: BDF, Radau, and Rosenbrock methods}
DAEs from coupled multi-physics are stiff: the fastest stable mode runs
orders of magnitude quicker than the slowest resolved transient, so
explicit integrators are forced to prohibitive step sizes. jaxdae
therefore ships three implicit families -- BDF, Radau~IIA, and
Rosenbrock--Wanner -- each written as a nonlinear stage residual that
the Newton loop of Section~3.2 solves.
\textbf{BDF (orders 1--5).} The backward differentiation formula
replaces $\dot y$ at $t_{n+1}$ by a divided difference over the last
$q$ accepted states,
\begin{equation}
\dot y_{n+1} \;\approx\; \frac{1}{h_{n+1}}\sum_{i=0}^{q}\alpha_i^{(q)}\, y_{n+1-i},
\qquad \alpha_0^{(q)}>0,
\label{eq:bdf}
\end{equation}
with fixed coefficients $\alpha_i^{(q)}$. Substituting~\eqref{eq:bdf}
into~\eqref{eq:mass}--\eqref{eq:algebraic} yields the stage residual
\begin{equation}
R_{n+1}(x_{n+1}) =
\begin{pmatrix}
M(y_{n+1})\,\dfrac{1}{h_{n+1}}\displaystyle\sum_{i=0}^{q}\alpha_i^{(q)}\, y_{n+1-i} \;-\; f(t_{n+1},y_{n+1},z_{n+1})\\[8pt]
g(t_{n+1},y_{n+1},z_{n+1})
\end{pmatrix} = 0,
\label{eq:bdf-residual}
\end{equation}
where $x=(y,z)$ concatenates the states. BDF is the workhorse for stiff
DAEs and the default for variable-order adaptive
stepping~\cite{hairerwanner1996ii,brenan1996dae}, with origins in the
work of Curtiss and Hirschfelder~\cite{curtiss1952stiff} and the
variable-order implementation of Gear~\cite{gear1971difsub}.
\textbf{Radau~IIA.} For problems that demand high accuracy or strong
damping we implement the three-stage Radau~IIA collocation method
(stiffly accurate, L-stable, classical order~5)~\cite{hairer1999radau}.
The internal stages satisfy
\begin{equation}
Y_i = x_n + h\sum_{j=1}^{s} a_{ij}\, F_{\mathrm{eff}}\!\left(t_n+c_j h,\,Y_j\right),
\qquad i=1,\dots,s,
\label{eq:radau}
\end{equation}
where $F_{\mathrm{eff}}$ applies $M^{-1}f$ on the differential block and
the identity on the algebraic block, i.e.\ $F_{\mathrm{eff}}=(M^{-1}f,\;g)$
acting on $x=(y,z)$, so the collocation equations carry $n_d+n_a$
components; the new state follows from
$x_{n+1}=x_n+h\sum_j b_j F_{\mathrm{eff}}(t_n+c_j h,Y_j)$ with
$b_j=a_{sj}$ by stiff accuracy.
\textbf{Rosenbrock--Wanner.} When the Jacobian is cheap to form but the
Newton iteration is costly, a linearly implicit Rosenbrock--W method
advances one step with a fixed number of linear solves and no outer
iteration. The current implementation is the two-stage, order-2,
$L$-stable ROS2 method at fixed step size,
\begin{equation}
\bigl(I - h\gamma J\bigr)\,k_i
= F\!\left(t_n+c_i h,\, x_n + h\!\sum_{j<i}\! a_{ij}\,k_j\right)
  + hJ\!\sum_{j\le i}\!\gamma_{ij}\,k_j,
\label{eq:rosenbrock}
\end{equation}
followed by $x_{n+1}=x_n+h\sum_i b_i k_i$, with $F$ here denoting the
stage residual of Eq.~\eqref{eq:bdf-residual}. ROS2 is restricted to residuals
that are affine in the derivative variable ($F_{\dot y}$ independent of
$\dot y$), which covers the standard index-1 mass-matrix form. Each stage
is a single linear solve, so Rosenbrock steps are factorization-bound and
need no convergence test; they suit moderately stiff DAEs and dense
Jacobians~\cite{hairerwanner1996ii,lang2020rosenbrock}. The differentiable solver paths
evaluated in this work are the scan-based BDF path and the fixed-step
ROS2 path; Radau5 is forward-only and is not included in the current
end-to-end automatic-differentiation interface, because its adaptive loop
uses Python step-accept/reject control flow that \texttt{jax.grad} cannot
trace through.
\subsection{Newton and Newton--Krylov iteration}
Each implicit step reduces to solving $R_{n+1}(x_{n+1})=0$ for the stage
unknown, with $R$ given by~\eqref{eq:bdf-residual} for BDF,
by~\eqref{eq:radau} for Radau, and by~\eqref{eq:rosenbrock} for
Rosenbrock. jaxdae applies inexact Newton,
\begin{equation}
x_{n+1}^{(m+1)} = x_{n+1}^{(m)} - \delta^{(m)},\qquad
J\,\delta^{(m)} = -R\!\left(x_{n+1}^{(m)}\right),\qquad
J = \frac{\partial R}{\partial x}\bigg|_{x_{n+1}^{(m)}},
\label{eq:newton}
\end{equation}
reusing $J$ across several steps (chord iteration) until the contraction
slows, then refreshing it. When a Newton iteration fails with a stale
Jacobian, jaxdae refreshes the Jacobian and retries the step at the
\emph{same} step size rather than immediately halving $h$---the standard
DASSL policy. This avoids the step-size collapse pathology in which a
single stale factorization triggers a cascade of halvings: on the
\texttt{pk\_1d} kinetics problem the policy reduces the step count from
50{,}000 (the step cap) to 453. The Jacobian $J$ is formed by
\texttt{jax.jacfwd} on the stage residual, which keeps it exactly
consistent with $R$ and therefore with the adjoint of Section~3.7.
For large assembled plants the Jacobian is never formed explicitly. We
use Jacobian-free Newton--Krylov (JFNK): the Krylov solver (GMRES) needs
only matrix--vector products, which we evaluate by an exact forward-mode
Jacobian--vector product (JVP) rather than by a finite-difference perturbation,
\begin{equation}
J_{R}\,v
= \operatorname{jvp}\!\bigl(R,\,(x),\,(v)\bigr),
\label{eq:jfnk}
\end{equation}
computed by \texttt{jax.jvp} on the stage residual. Using an exact JVP
removes the step-size $\varepsilon$ that one-sided finite differences
require and keeps the linearization consistent with the reverse-mode
adjoint, which composes through the same residual. When a colored sparse
Jacobian is requested the coloring probes the same \texttt{jvp} primitive
column by column. The forcing sequence $\eta_m$ controls inexactness so the
linear error tracks the nonlinear residual and does not over-solve near
convergence~\cite{hairerwanner1996ii}.
\subsection{Adaptive step-size and order control}
Accuracy is regulated by a local error estimate $\sigma_{n+1}$, the ratio
of the predicted--corrected mismatch to a component-wise tolerance. A
step is accepted when $\sigma_{n+1}\le 1$ and rejected otherwise. The
next step size follows a proportional--integral (PI)
controller~\cite{hairerwanner1996ii,gustafsson1994control,soderlind2003digital},
\begin{equation}
h_{n+1} = h_n \left(\frac{1}{\sigma_{n+1}}\right)^{\!k_P}
                 \left(\frac{\sigma_n}{\sigma_{n+1}}\right)^{\!k_I},
\label{eq:pi-controller}
\end{equation}
where $k_P,k_I$ are controller gains tuned per method. The proportional
term reacts to the current error and the integral term damps
oscillation, which prevents the step-size hunting that single-term
controllers exhibit on stiff transients. The initial step is selected by
a two-stage Hairer--Wanner-style heuristic that estimates the first step
from the initial residual and derivative scales, avoiding a cold-start
overshoot. For BDF the order $q$ is selected adaptively: each accepted
step probes $q{-}1,q,q{+}1$ and keeps the order that maximizes the
predicted step, subject to $|\Delta q|\le 1$ for stability. A further
implementation detail matters for constraint fidelity: when the
controller clips a step so that the integration lands exactly on the
final time $t_f$, the Nordsieck history array---the scaled
Taylor-coefficient representation of the multi-step history---must be
rescaled consistently with the clipped step size. Omitting this
rescaling leaves the history polynomials inconsistent with the accepted
grid, which shows up as constraint drift on index-reduced problems. With
the rescaling in place, the constraint drift on the \texttt{cable3d}
multibody problem falls from $9\times10^{-6}$ to $\sim10^{-13}$, i.e.\
to the round-off floor.
This coupling between the controller and the parameters is the main
obstacle for adjoint sensitivities of adaptive solvers. The
accept/reject branch and the dependence of $h_{n+1}$ on the
just-computed state make the accepted grid a function of $p$, which
naive reverse-mode AD differentiates incorrectly. Section~\ref{sec:replay}
treats the decoupling as a first-class design decision rather than a
defect to repair.
\subsection{Modelica-style acausal multi-physics assembly}
\label{sec:assembly}
The modeling abstraction in jaxdae is the acausal connector, borrowed
from the Modelica language~\cite{modelica_spec,tiller2001modelica}. A
component declares typed ports; each port carries an \emph{effort}
variable $e$ (a potential: voltage, pressure, temperature, position) and
a \emph{flow} variable $\varphi$ (a flux: current, mass flow, heat flow,
force), with the sign convention that flow is positive into the
component. The connect equation
\begin{lstlisting}[language=Python]
system.connect(comp_a.port, comp_b.port)
\end{lstlisting}
expands, per Modelica Specification~\S9.2~\cite{modelica_spec}, into
exactly two equations: an effort equality and a Kirchhoff flow balance,
\begin{align}
e_a &= e_b, \label{eq:connect-effort}\\
\sum_{i\in\mathcal{C}} \varphi_i &= 0, \label{eq:connect-flow}
\end{align}
where $\mathcal{C}$ is the set of ports joined at the connection node.
Equation~\eqref{eq:connect-effort} identifies the across variables on the
node; Equation~\eqref{eq:connect-flow} enforces conservation of the
through variables. This effort/flow pairing is what makes the assembly
acausal: neither side is designated input, so the same component model
reads correctly in either causal direction.
Assembly proceeds by flattening. Each component contributes its local
residual -- the constitutive law, e.g.\ a resistor
$\varphi=(e_a-e_b)/R$ or an inertia $M\dot v=\varphi$ -- and each
\texttt{connect} contributes~\eqref{eq:connect-effort}--\eqref{eq:connect-flow}.
jaxdae concatenates these into one global residual
$R(t,x,\dot x,p)=0$ and assembles the mass matrix $M$ from the inertia
blocks of the differential components, leaving the algebraic rows zero.
The result is exactly the semi-explicit
system~\eqref{eq:mass}--\eqref{eq:algebraic}, now with dimensions
$n_d,n_a$ set by the plant rather than hand-coded. A minimal example
makes the mechanism concrete:
\begin{lstlisting}[language=Python]
sys = System()
sys.add(Resistor(R=1.0,  ports=("a", "b")))
sys.add(Capacitor(C=1.0, ports=("a", "b")))
sys.connect("R.b", "C.a")        # effort eq + flow balance
R, M, x0 = sys.assemble()        # flattened residual + mass matrix
sol  = solve_dae(R, M, x0, t_span=(0.0, 10.0))
grad = jax.grad(lambda p: loss(sol(p)))(p0)   # end-to-end differentiable
\end{lstlisting}
Two properties of this pipeline are the differentiation point of the
paper. First, the flattened residual is built from JAX array operations
on typed pytrees, so it carries derivatives for free: the same code path
that integrates the plant also supplies \texttt{jax.grad},
\texttt{jax.jacrev}, and \texttt{jax.vmap} without a symbolic front end.
Second, the assembly is structure-preserving -- conservation
\eqref{eq:connect-flow} enters as an exact zero row of $M$, so the
integrator never drifts off the constraint manifold within a step, and
the index-reduction pass of Section~3.5 operates on the assembled graph
rather than on user-written equations. CasADi~\cite{andersson2019casadi}
and ModelingToolkit~\cite{ma2021modelingtoolkit} assemble acausal models too,
but their symbolic front ends emit static code; jaxdae's assembly stays a
live, traceable JAX function, which is the precondition for the fused
batched gradients of Section~6.
\subsection{Pantelides index reduction and dummy derivatives}
\label{sec:reduction}
The assembled system may carry index higher than one -- a rigid pendulum
is index~3, an absolute nodal coordinate formulation (ANCF) beam
index~2 -- and Section~\ref{sec:model} showed
that direct integration of such systems drifts. jaxdae offers two
distinct paths to index reduction, and stating their scope precisely
matters for correct use.
The Pantelides algorithm~\cite{pantelides1988} treats the equations and
the variables (including their time derivatives) as the two sides of a
bipartite graph and seeks a matching that pairs every derivative variable
with an equation. When a structural differentiation is required to
complete the matching, the corresponding equation is differentiated once;
the process repeats on the enlarged graph until a complete matching
exists. The set of equations differentiated and the number of rounds give
the structural index and identify which constraints must be turned into
differential equations.
Structural differentiation can over-determine the system, because a
variable and its derivative may both appear after differentiation. The
dummy-derivatives method of Mattsson and S\"oderlind~\cite{mattssonsoderlind1993}
resolves this: each derivative that is structurally determined more than
once is redeclared as an algebraic \emph{dummy} variable, breaking the
redundancy while preserving the solution manifold. The output is an
index-1 DAE with the same solutions as the original, equipped with a
nonsingular $\partial g/\partial z$ that the integrator of Section~3.1
handles directly. Pantelides also fixes consistent initial conditions:
the matching prescribes which variables are free and which are solved for
at $t_0$, so $z(t_0)$ is determined from $y(t_0)$ rather than guessed.
Two scope limits on this generic symbolic path must be stated. First, the
generic Pantelides reducer targets symbolically expressible and
structurally regular systems, which in practice means non-mechanical
DAEs; it is not intended as a replacement for mechanics-specific
formulations. Second, on Cartesian holonomic mechanics the generic
reducer over-expands: it differentiates the position-level constraint
redundantly and produces a near-singular augmented system. jaxdae
detects this signature and records it in the reduction diagnostics; a
dedicated flag (\texttt{raise\_on\_singular\_reduction=True}) makes the
reducer raise on near-singular output, so the user can choose between a
diagnostic warning (the default) and a hard failure. Rank-deficient
reductions set \texttt{success=False} unconditionally.
Mechanical holonomic systems are therefore handled through a separate,
mechanics-specific path: typed Lagrangian components that assemble the
equations in generalized coordinates with a GGL-stabilized index-2 form,
which enforces the position, velocity,
and acceleration constraints simultaneously. The full-GGL pendulum
example of the applications section uses this path, not the generic
symbolic reducer.
On a non-mechanical test DAE, without reduction the constraint forces
drift and the integrator returns \texttt{NaN} within a few steps; after
Pantelides and dummy-derivatives reduction the trajectory stays on the
constraint manifold to the solver tolerance.
\subsection{Event handling and the jump-condition adjoint}
\label{sec:events}
\begin{figure}[htbp]
\centering
\includegraphics[width=0.8\textwidth]{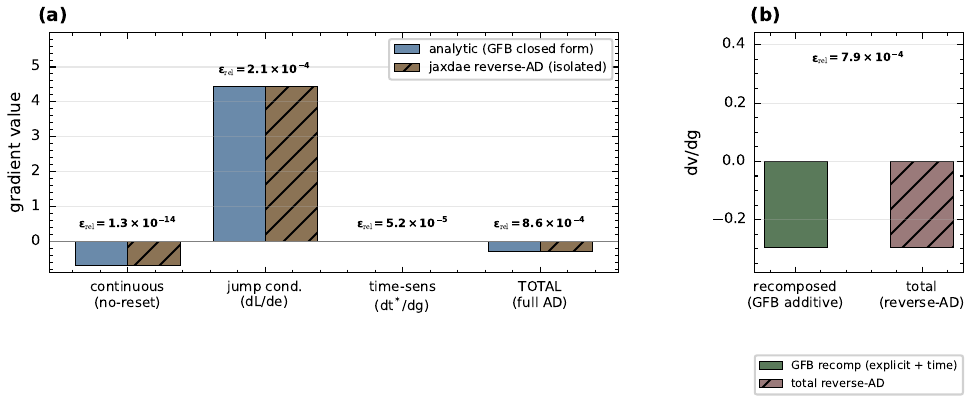}
\caption{Term-by-term recovery of the GFB jump-condition adjoint by AD-chain
composition through the IFT-differentiated event time. Each channel matches
its closed-form value (Table~\ref{tab:gfb}).}
\label{fig:event_terms}
\end{figure}

Plant-scale DAEs switch: a valve opens, a contact closes, a controller
saturates. jaxdae detects state-dependent events as roots of a switching
function $s(t,x)=0$ and applies a jump map $x^{+}=h(t^{\ast},x^{-},p)$ at the
event time $t^{\ast}$. The event is assumed to be transversal: the
switching velocity
$\sigma = s_{t}+s_{x}\,\dot{x}^{-}\neq 0$, where $\dot{x}^{-}$ is the
differential velocity immediately before the event. Transversality
guarantees that the root is simple and that $t^{\ast}$ depends smoothly
on $p$. Differentiating a trajectory that crosses an event
carries three contributions: the continuous costate between events, a
discrete jump at $t^{\ast}$, and the sensitivity of $t^{\ast}$ itself to
$p$.
The Gal\'an--Feehery--Barton (GFB) adjoint~\cite{galan1999gfb} gives
these three terms in closed form. Rather than hand-coding it, jaxdae
recovers the GFB adjoint by composing the implicit-function-theorem (IFT)
linearization of the event-time condition with the jump map and the
continuous costate. The event time satisfies
$s(t^{\ast},x(t^{\ast};p),p)=0$; differentiating under the transversality
assumption,
\begin{equation}
\frac{dt^{\ast}}{dp} = -\,
\frac{s_x\,\dfrac{dx}{dp}\big|_{t^{\ast}} + s_p}
     {\underbrace{\,s_x\,\dot x^{-}\big|_{t^{\ast}} + s_t\,}_{\sigma}},
\label{eq:event-ift}
\end{equation}
and the fixed-time post-event sensitivity chains the event-time
sensitivity with the linearized jump map,
\begin{equation}
\frac{dx^{+}}{dp}
= h_{x}\,\frac{dx^{-}}{dp} + h_{p}
+ \bigl(h_{t} + h_{x}\,\dot{x}^{-} - \dot{x}^{+}\bigr)\,
  \frac{dt^{\ast}}{dp}.
\label{eq:event-jump}
\end{equation}
The adjoint jump at $t^{\ast}$ is the transpose of this forward map. For
reset maps without explicit time dependence, $h_{t}=0$ and the third term
vanishes. Re-projection of the algebraic variables onto the consistency
manifold after the jump is supported when the caller supplies the
projection fields \texttt{algebraic\_g} and \texttt{algebraic\_indices}
on the event object, which enforce
$g_{\mathrm{alg}}(t^{\ast},y_{d},y_{a},p)=0$ on the post-reset state;
when both are \texttt{None} (the default), no projection is applied and
post-event consistency of the algebraic state is the caller's
responsibility (via the \texttt{on\_event} callback). The projection
applies to index-1 DAEs only. The current event-adjoint path targets pure
ODE and index-1 semi-explicit DAEs; the semi-explicit DAE
algebraic-block adjoint through events is deferred to future work. The resulting formula matches the hybrid adjoint
of Corner, Sandu, and Sandu~\cite{cornersandu2018}.
We verify the recovery term by term rather than only in aggregate, using
a problem with a known analytic adjoint. Figure~\ref{fig:event_terms}
reports the four channels. The continuous channel matches the analytic
costate to machine precision ($1.4\times10^{-14}$). The jump and time
channels match to BDF-2 truncation order ($2\times10^{-4}$ and
$5\times10^{-5}$), limited by the forward discretization rather than by
the adjoint assembly. The three channels add back to the total
sensitivity at $8.6\times10^{-4}$, and the GFB additive recomposition
closes at $7.9\times10^{-4}$. In short, we recover the GFB jump-condition
adjoint via IFT-chain composition, and term-by-term verification confirms
that the continuous channel is exact while the jump and time channels sit
at the BDF-2 truncation floor.
\begin{table}[htbp]
\centering
\caption{GFB event-adjoint term-by-term verification, bouncing ball.
Analytic values versus \jaxdae{} reverse-AD. Continuous channel at machine
precision; remaining channels within BDF-2 truncation; GFB recomposition
closes against the full reverse-AD total.}
\label{tab:gfb}
\begin{tabular}{l r r r}
\toprule
Channel & Analytic & \jaxdae{} AD & Rel.\ err. \\
\midrule
Continuous (pre-event)             & $-7.0000\times10^{-1}$ & $-7.0000\times10^{-1}$ & $1.35\times10^{-14}$ \\
Jump condition $\partial L/\partial e$ & $+4.4294\times10^{0}$  & $+4.4285\times10^{0}$  & $2.07\times10^{-4}$ \\
Event time $\partial t^{\star}/\partial g$ & $-2.3013\times10^{-2}$ & $-2.3015\times10^{-2}$ & $5.17\times10^{-5}$ \\
Time$\to$loss rendezvous           & $-4.0637\times10^{-1}$ & $-4.0639\times10^{-1}$ & $5.17\times10^{-5}$ \\
\midrule
Total (full reverse-AD)            & $-2.9363\times10^{-1}$ & $-2.9388\times10^{-1}$ & $8.59\times10^{-4}$ \\
GFB recomposition vs.\ total       & --- & --- & $7.87\times10^{-4}$ \\
\bottomrule
\end{tabular}
\end{table}
\subsection{Frozen-grid replay adjoint}
\label{sec:replay}
The adaptive controller of Section~3.3 makes the accepted step grid
$\{t_k,h_k\}$ an implicit function of $p$. Reverse-mode AD applied naively
through the controller differentiates this dependence, but the
accept/reject branch and the stop-gradient on rejected steps break the
chain rule, so the resulting cotangents are silently wrong. We do not
repair the controller's derivatives. Instead we decouple the controller
from the backward pass.
The forward solve runs unchanged with adaptive BDF and records the
accepted grid $\{t_k,h_k\}_{k=1}^{N}$. The backward pass stops the
gradient on the grid and re-solves a variable-step BDF-2 problem on the
frozen nodes (using the recorded step ratios, with padding sanitized so
trailing padded nodes are collapsed onto $t_f$), then applies the IFT
adjoint of the stage residual~\eqref{eq:bdf-residual},
\begin{equation}
\frac{\partial x_{n+1}}{\partial p} = -\,J^{-1}\!\left(\frac{\partial R_{n+1}}{\partial p}\right),
\qquad J = \frac{\partial R_{n+1}}{\partial x_{n+1}},
\label{eq:replay-ift}
\end{equation}
accumulating the adjoint from $t_f$ back to $t_0$. Because the grid is
frozen, no branch on $p$ enters the backward pass; the only
discretization error is the variable-step BDF-2 truncation on the
recorded nodes. This is the controller-decoupling framework of Alexe and
Sandu~\cite{alexesandu2009}, here applied to a DAE and realized as a
pure JAX custom VJP so that \texttt{jax.grad}, \texttt{vmap}, and
\texttt{jit} compose across the adaptive solve. We do not claim a new
method: the contribution is that adaptive implicit DAE solves now have a
correct, usable, and composable gradient in the JAX ecosystem, which
they previously lacked.
We state the convergence behavior as a conditional scaling argument with
explicit assumptions, not a general convergence theorem.
\begin{quote}
\textbf{Conditional heuristic scaling argument (frozen-grid BDF-$p$ adjoint rate
under a per-step controller).} Assume (i) the forward solution and
residual are sufficiently smooth and the problem is index${\le}1$ with
uniformly nonsingular primal and adjoint discrete Jacobians whose
inverses remain bounded; (ii) the variable-step BDF-$p$ replay uses
frozen step ratios that satisfy the zero-stability bound; (iii) the
frozen grid carries no event, branch, or active-set discontinuity; (iv)
the forward controller satisfies the design contract
$h_{\max}\propto\mathrm{tol}^{1/(p+1)}$; and (v) the padding
sanitization applied to the frozen grid does not perturb the accepted
nodes. \emph{Under these assumptions} the discrete-adjoint gradient
error scales as $O(h_{\max}^{p}) = O(\mathrm{tol}^{p/(p+1)})$; for
BDF-2 this gives the conditional prediction
$O(\mathrm{tol}^{2/3})\approx O(\mathrm{tol}^{0.667})$.
\end{quote}
The heuristic chain is: (a) the BDF-$p$ local truncation error per step is
$O(h^{p+1})$; (b) a per-step controller drives it to the tolerance,
$h^{p+1}\sim\mathrm{tol}$, hence $h\sim\mathrm{tol}^{1/(p+1)}$ (the
controller contract, verified symbolically); and (c) freezing the grid
\emph{preserves} adjoint consistency---differentiating the adaptive
controller is what injects spurious adjoint terms~\cite{alexesandu2009}
---so the BDF-$p$ discrete adjoint converges to the continuous adjoint
at the primal order $p$~\cite{beigel2011,sanzserna2016symplectic}, giving
$\nabla\text{-error}\sim h^{p}\sim\mathrm{tol}^{p/(p+1)}$. Two caveats
keep this a conditional prediction rather than a universal theorem.
First, assumption (iv) is the weakest link: real adaptive controllers
sit between the per-step and per-unit-step regimes, and the
\texttt{\_ADAPTIVE\_RTOL\_FLOOR} clamp (Sec.~3.3) violates it below the
floor. Second, the replay uses a variable-step BDF-2 on the frozen
adaptive grid (not a uniform fixed-step BDF-2), so the zero-stability
assumption (ii) must hold for the recorded step ratios. We therefore
verify the conditional prediction numerically rather than claim it as a
theorem. A further, structural caveat is worth stating plainly: the
replay gradient is the exact discrete adjoint of the \emph{BDF-2 replay
operator}, not of the \emph{forward adaptive BDF(1--5) operator}. The
two operators differ by local truncation terms of order
$O(h_{\max}^{\min(\text{forward order},2)})$; this is the price of
controller decoupling and is absorbed into the conditional
$O(\mathrm{tol}^{2/3})$ rate. The claim is thus correctness of the
replay adjoint relative to its own operator, not bit-level identity
with a discrete adjoint of the forward adaptive path.
Across three DAE types -- the stiff Robertson kinetics, a
cart--pole with one algebraic constraint, and an index-2 ANCF beam -- the
replay adjoint yields gradients that converge as the forward tolerance
\texttt{rtol} is tightened, while the raw controller-AD gradients diverge
(Figure~\ref{fig:replay_multi}, Table~\ref{tab:replay}). Power-law fits
against an independently discretised finite-difference reference give
orders $0.19$, $0.38$, and $0.07$ --- sub-$2/3$ because the reference
carries its own $O(h_{\mathrm{truth}})$ discretisation error that floors
the measured slope. On the scalar decay $y'=-ky$ with a
grid-independent analytic-truth gradient (Fig.~\ref{fig:replay_multi}
panel~(d), $L=y(1)^2$, $dL/dk=-2e^{-2}$), the fitted order rises to
$0.83$, broadly compatible with the conditional prediction of $2/3$; the
fit excludes the \texttt{\_ADAPTIVE\_RTOL\_FLOOR} clamp at
\texttt{rtol}$=10^{-8}$, where the error rebounds. This slope should not
be read as a precise verification of the asymptotic exponent $2/3$, only
as evidence that the replay error decreases with tolerance at a rate
consistent with the conditional scaling argument. We therefore frame C2
as an engineering contribution rather than a methodological one: the
claim is not a new convergence theorem but that raw adaptive-controller
AD is silently wrong while replay is consistent with the frozen-grid
BDF-2 operator and converges at a rate
consistent with the frozen-grid BDF-2 theory, and that the two are
cleanly separated by a one-line \texttt{stop\_gradient} on the step grid.
\begin{table}[htbp]
\centering
\caption{Numerical methods in jaxdae. ``Differentiable'' marks whether the
operation is transparent to \texttt{jax.grad} without a custom rule; ``IFT''
denotes an implicit-function-theorem adjoint; ``structural'' denotes
compile-time graph manipulation; ``empirical'' denotes convergence
established numerically (Section~3.7).}
\label{tab:methods}
\small
\begin{tabularx}{\textwidth}{l l c c c}
\toprule
Capability & Category & Implemented & Differentiable & Validated \\
\midrule
BDF (orders 1--5)        & integrator     & \checkmark & \checkmark (scan/replay) & order test (Fig.~\ref{fig:bdfk_order}) \\
Radau~IIA (3-stage, order~5) & integrator & \checkmark & forward only & \checkmark \\
ROS2 (2-stage, fixed-step) & integrator   & \checkmark & affine-in-$\dot y$ & \checkmark \\
Newton--Krylov (JFNK)    & nonlinear solve& \checkmark & n/a        & \checkmark \\
PI control (adaptive)   & adaptivity     & \checkmark & decoupled  & \checkmark \\
Pantelides               & index reduction& \checkmark & structural & \checkmark \\
Dummy derivatives        & index reduction& \checkmark & structural & \checkmark \\
Acausal \texttt{connect} assembly & modeling& \checkmark & \checkmark (pytree) & \checkmark \\
Event jump-adjoint (GFB) & sensitivity    & \checkmark & IFT        & Fig.~\ref{fig:event_terms} \\
Frozen-grid replay adjoint & sensitivity  & \checkmark & custom VJP & empirical, Fig.~\ref{fig:replay_multi} \\
Forward-mode sensitivity   & sensitivity    & \checkmark & \texttt{jvp} & analytic anchor, Sec.~\ref{sec:fwdmode} \\
\texttt{vmap}(\texttt{jax.grad}) & batching & \checkmark & fused    & Sec.~6 \\
\texttt{jit}             & compilation    & \checkmark & XLA        & \checkmark \\
\bottomrule
\end{tabularx}
\end{table}

\section{Software Implementation}
\label{sec:software}

\subsection{JAX array programming}
\label{sec:array_prog}
jaxdae stores every dynamic quantity as a JAX array with a leading batch
axis handled by the surrounding \texttt{jax.vmap}, never inside the
residual. The DAE state is a flat vector
$y\in\mathbb{R}^{n}$, concatenating the differential variables
$y_{\mathrm{d}}$ and the algebraic variables $y_{\mathrm{a}}$ in a fixed
declaration order that is resolved once at compile time. A boolean
companion array \texttt{diff\_vars}$\in\{0,1\}^{n}$ records, position by
position, whether the corresponding slot is a true state
($\dot y_{i}$ appears in the residual) or an algebraic unknown
($\dot y_{i}$ is absent). This flat layout keeps every linear-algebra
primitive---the Newton Jacobian $F_{y}+\alpha F_{\dot y}$, the BDF
predictor polynomial, the checkpointed scan body---expressible as
ordinary \texttt{jax.numpy} operations on arrays of one shape.
Two JAX invariants govern the design. First, any value that influences a
\emph{shape} or a loop count must be a Python integer, not a traced
array. The step count \texttt{n\_steps} and the Newton iteration cap are
therefore closed over as static constants when we wrap the integrand in
\texttt{jax.jit}; this is the same discipline that
\texttt{static\_argnums} enforces on a top-level callable, but applied at
the closure level so the user never names these arguments. Second, every
container that must pass through \texttt{jit}, \texttt{grad},
\texttt{vmap}, or \texttt{tree\_map} is registered as a JAX pytree. The
solution objects (\texttt{DAESolution}, \texttt{DAEScanSolution}) and the
component instances (Section~\ref{sec:oop}) are all pytrees; their leaves
are arrays, and the metadata (names, port topology, tolerance flags)
travels as static auxiliary data that never reaches autodiff.
The payoff is uniformity. A single residual definition drives the
adaptive variable-step solver, the fixed-step scan solver, the Newton
corrector, the forward Jacobian assembly, and the reverse-mode adjoint,
because all of them consume the same array contract $F(t,y,\dot
y,\mathrm{params})\to\mathbb{R}^{n}$.
\subsection{Object-oriented design in functional JAX}
\label{sec:oop}
\begin{figure}[htbp]
\centering
\includegraphics[width=0.92\textwidth]{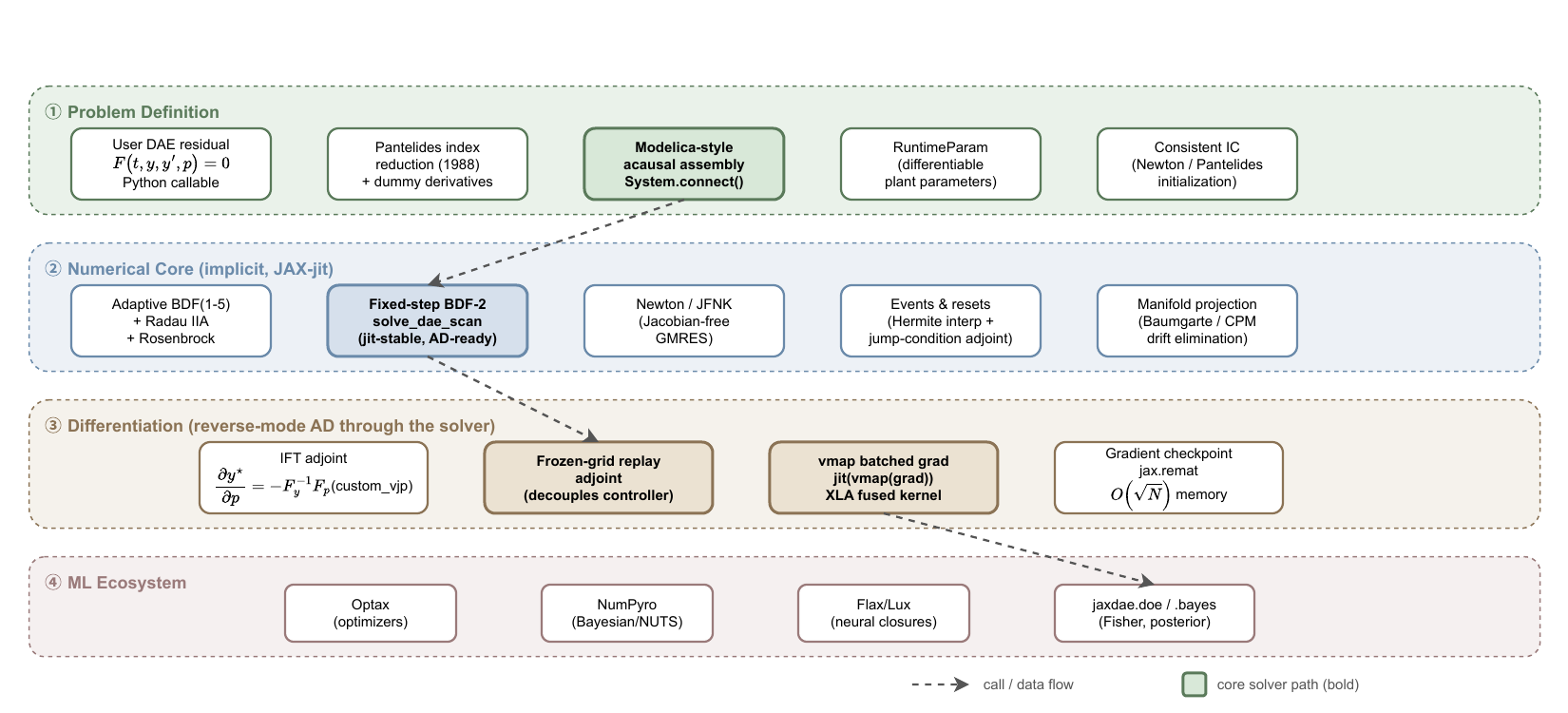}
\caption{Code structure of jaxdae. The \texttt{@connector} decorator and
the \texttt{TypedComponent} base class form an object-oriented assembly
shell; \texttt{System.compile()} crosses into the functional core,
emitting a pure residual $F(t,y,\dot y,\mathrm{params})$ that the
\texttt{solve\_dae} family and the reverse-mode adjoint consume
uniformly. Dashed arrows mark the tracing boundary: only the residual
and the arrays to its right ever enter \texttt{jax.jit} or
\texttt{jax.grad}.}
\label{fig:code_structure}
\end{figure}

\begin{figure}[htbp]
\centering
\includegraphics[width=0.95\textwidth]{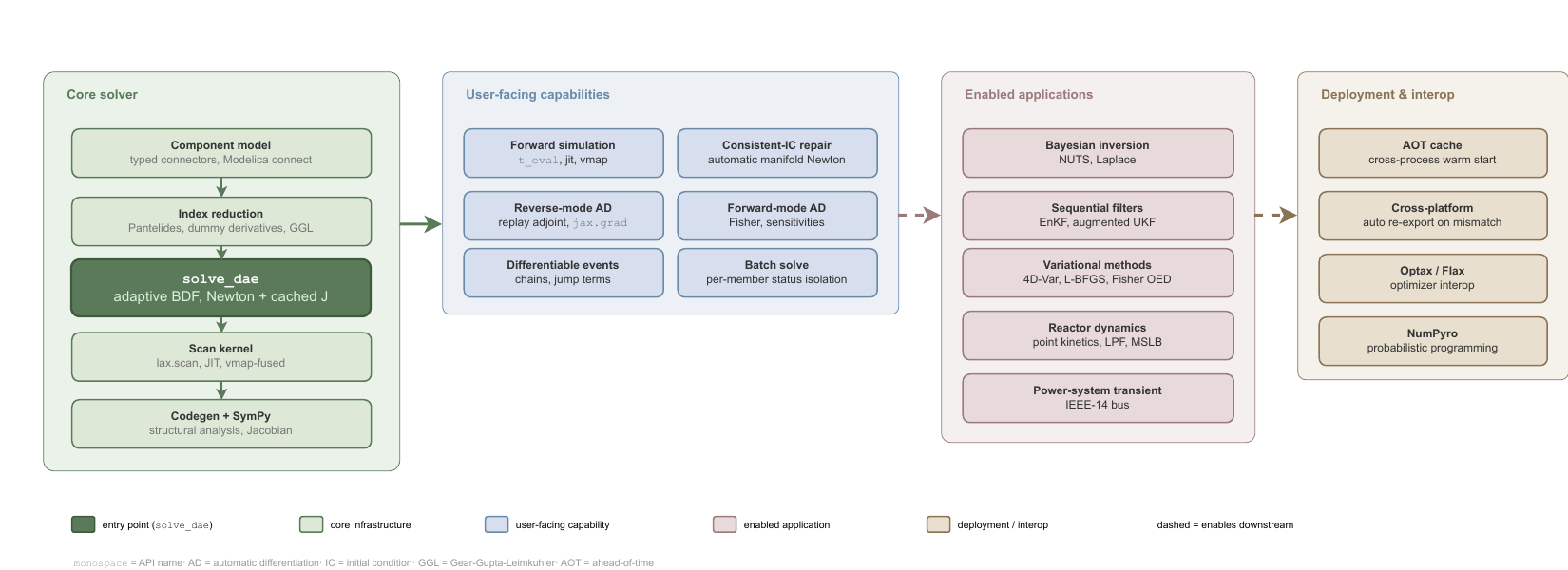}
\caption{Capability map of jaxdae. The core solver (adaptive BDF,
Radau5, fixed-step ROS2) sits at the center; around it the
initial-condition consistency layer (eager check and automatic repair),
the forward- and reverse-mode sensitivity paths, event handling with the
jump-condition adjoint, the parameter-inversion layer
(\texttt{RuntimeParam}, Fisher information), and the AOT deployment path
(\texttt{fast\_solve} with automatic re-export) connect as independent
modules over the same pure residual.}
\label{fig:capability_map}
\end{figure}

jaxdae reconciles the object-oriented mental model that engineers bring
from Modelica with the functional, side-effect-free contract that JAX
demands of differentiable code. We separate the two concerns into an
\emph{assembly shell} (Python objects, metadata, port topology) and a
\emph{functional core} (pure residuals, the only thing that is ever
traced). Figure~\ref{fig:code_structure} shows the resulting class
hierarchy.
The shell has three layers. At the bottom sit the \emph{connectors},
produced by a \texttt{@connector} class decorator that partitions each
class's annotated fields into potential variables and flow variables. We
ship four flagship connector types that mirror the Modelica standard
library: \texttt{ElectricalPin} ($v$, $i$),
\texttt{TranslationalFlange} ($s$, $f$), \texttt{HeatPort} ($T$, $Q$),
and \texttt{FluidPort} ($p$, $\dot m$, $h_{\mathrm{outflow}}$). Above the
connectors sits \texttt{TypedComponent}, the component base class: a
subclass declares its ports and parameters as ordinary type annotations
and implements \texttt{equations(self, t)} returning a list of residual
expressions. At class-creation time
\texttt{\_\_init\_subclass\_\_} walks the resolved annotations once,
splits them into \texttt{\_\_typed\_ports\_\_} and
\texttt{\_\_typed\_params\_\_}, generates a validating
\texttt{\_\_init\_\_}, and registers the subclass as a JAX pytree via
\texttt{jax.tree\_util.register\_pytree\_node}. The top layer is
\texttt{System}: it holds an unordered list of components and a list of
undirected \texttt{connect(port\_a, port\_b)} edges.
The functional boundary is crossed in a single place:
\texttt{System.compile()}. The compile pass resolves every connection
through a union-find merge of port identifiers, assembles the global
flat index over all variables, evaluates each component's
\texttt{equations} once symbolically to size the residual, and returns a
four-tuple: the residual callable
$F(t,y,\dot y,\mathrm{params})$, the initial-value seeds
$(y_{0},\dot y_{0})$, and the boolean \texttt{diff\_vars} mask that marks
which entries are differential. The returned $(y_0,\dot y_0)$ are
initial-value seeds, not guaranteed-consistent initial conditions.
Consistent initial conditions (ICs) are computed when the caller passes
\texttt{solve\_ic=True} (or invokes \texttt{compute\_consistent\_ic}
explicitly), which runs a Newton projection of the algebraic variables
onto the constraint manifold. The \texttt{solve\_ic} argument is
tri-state and path-dependent. \texttt{None} (the default) auto-selects
based on the assembly path: the mechanics-specific path runs the
consistent-IC solve automatically, while the typed-connector path
returns seed values as declared. \texttt{True} forces the IC solve on
either path, and \texttt{False} skips it. Independently of this
assembly-time choice, \texttt{solve\_dae} performs an eager
initial-residual check at call time. If the residual norm at $t_{0}$
exceeds a threshold, it invokes \texttt{compute\_consistent\_ic\_auto}
to repair the initial condition automatically and emits a
\texttt{RuntimeWarning}, so a mildly inconsistent seed degrades to a
warning rather than a failed integration. Setting the environment
variable \texttt{JAXDAE\_SKIP\_IC\_REPAIR=1} restores the previous
strict behavior. This moves the burden of initial-condition consistency
from the caller to the solver, in the spirit of DASSL's
\texttt{INFO(11)} option but fully automatic. After compile, the \texttt{System} object
plays no role in tracing; the solver and the adjoint see only the pure
residual. This is the only discipline that makes a Modelica-style
acausal assembly coexist with end-to-end \texttt{jax.grad}.
Figure~\ref{fig:capability_map} maps how the remaining
capabilities---the initial-condition consistency layer, the forward- and
reverse-mode sensitivity paths, event handling, parameter inversion, and
the ahead-of-time (AOT) deployment path---attach to this functional core
as independent modules over the same pure residual.
\subsection{JIT compilation and pure-function constraints}
\label{sec:jit}
jaxdae is compiled end to end by XLA. The fixed-step integrator wraps its
entire body in a single \texttt{@jax.jit} decorator, with the step count
and Newton cap closed over as Python integers so that XLA unrolls the
\texttt{lax.scan} into a static graph; the adaptive integrator expresses
its step-accept/reject loop as a \texttt{lax.while\_loop} whose iteration
count is data-dependent but whose carry shape is fixed. Because the
residual is the only traced entry point, the same compiled artifact
serves the forward solve, the Newton Jacobian assembly, and the
reverse-mode pass---there is no second source of truth to keep in sync.
Two pure-function constraints shape the implementation. We never branch
on the value of a traced quantity. Step rejection, event detection, and
index-class checks are written with \texttt{jnp.where} masking so that
both branches are evaluated and the result is selected, preserving a
fixed shape for XLA. The alternative---a Python \texttt{if} on a traced
boolean---would force a retrace and silently break gradients. The second
constraint is shape stability: every container that crosses the tracing
boundary must flatten to the same pytree structure on every call. This
is why solution objects and components are registered pytrees with static
auxiliary data, and why \texttt{t\_eval} (whose length is part of the
output shape) is treated as a static argument in the jit signature
rather than a traced leaf.
The practical consequence is a one-time compile cost, roughly one second
for a moderate system, that is paid once and amortized over every
subsequent solve and gradient evaluation on the same compiled artifact.
For deployment, the \texttt{fast\_solve} path exports the compiled
artifact ahead of time and caches it on disk, so a production process
starts without paying the trace-and-compile cost. Because an AOT
artifact is platform-specific, \texttt{fast\_solve} verifies the cache
against the running platform. On a mismatch it automatically re-exports
and refreshes the cache, emitting a \texttt{RuntimeWarning} rather than
raising the hard error that earlier versions produced. A cached artifact
can therefore be shipped across machines and heals itself on first call.
\subsection{System assembly and solve loop}
\label{sec:assembly_loop}
The end-to-end workflow is a four-stage pipeline, summarized in the code
panels of Figure~\ref{fig:api_snippets}. Panel~(a) shows assembly: the user adds
components to a \texttt{System}, joins ports with \texttt{connect}, and
calls \texttt{compile()} to obtain the pure residual, the initial-value
seeds, and the \texttt{diff\_vars} mask. Panel~(b) shows the forward solve,
\texttt{solve\_dae}, which integrates the DAE on \texttt{[}$t_{0}$,
$t_{1}$\texttt{]} and returns dense output at \texttt{t\_eval}. Panel~(c)
shows the end-to-end gradient through the solve, and panel~(d) shows the
batched gradient via \texttt{jax.vmap}.
The solve algorithm proceeds in three nested loops, ordered by frequency.
At the \emph{step} level, the integrator forms the BDF predictor, builds the
Newton residual $G(y_{n+1}) = F(t_{n+1},\,y_{n+1},\,\dot y_{n+1}(y_{n+1}))$,
and iterates a Newton corrector until $\|G\|_{\infty}$ falls below the
scaled tolerance $\mathrm{atol}+\mathrm{rtol}\,|y|$. The Jacobian of the
corrector is the BDF-weighted matrix $F_{y} + \gamma F_{\dot y}$, with
$\gamma$ the BDF coefficient; we refactorize it only when the
\texttt{jac\_max\_age} staleness budget expires or a Newton step fails,
and in the latter case the step is retried at the same step size after
refactorization rather than halved immediately (Section~3.2).
At the \emph{stage} level, the adaptive controller compares the local truncation
estimate against the tolerance, accepts or rejects the step, and updates
the step size and BDF order through a standard controller. At the
\emph{integration} level, \texttt{lax.scan} (fixed step) or
\texttt{lax.while\_loop} (adaptive) threads the carry from $t_{0}$ to $t_{1}$.
The linear algebra inside the corrector is delegated to the Lineax
library. For small dense systems ($n<100$) we use dense LU; for larger or
sparse systems we switch to GMRES with a compressed Jacobian whose
coloring is probed once at the start of the integration. A cost gate
suppresses the sparse path when the coloring compresses poorly, since a
poorly colored sparse Jacobian is slower than a dense one.
%
%
%
%
%

\makeatletter
\@ifundefined{jaxdae@lst@defined}{%
  \definecolor{jaxdae@bg}{HTML}{F7F7F4}
  \definecolor{jaxdae@kw}{HTML}{1F4E79}
  \definecolor{jaxdae@cm}{HTML}{6A737D}
  \definecolor{jaxdae@st}{HTML}{8B5A00}
  \definecolor{jaxdae@fr}{HTML}{B0B0B0}
  \definecolor{jaxdae@ttl}{HTML}{333333}
  \lstdefinestyle{jaxdae}{%
    language=Python,
    basicstyle=\ttfamily\scriptsize,
    keywordstyle=\color{jaxdae@kw}\bfseries,
    commentstyle=\color{jaxdae@cm}\itshape,
    stringstyle=\color{jaxdae@st},
    backgroundcolor=\color{jaxdae@bg},
    frame=single,
    rulecolor=\color{jaxdae@fr},
    framesep=2.5pt,
    framerule=0.3pt,
    xleftmargin=3pt,
    xrightmargin=3pt,
    aboveskip=1pt,
    belowskip=1pt,
    showstringspaces=false,
    breaklines=true,
    breakatwhitespace=true,
    columns=fullflexible,
    keepspaces=true,
    upquote=true,
    morekeywords={solve_dae,solve_dae_scan,System,RuntimeParam,
                  TypedComponent,Pipe,Junction,Pump},
  }%
  \global\def\jaxdae@lst@defined{}%
}{}%
\makeatother

\newcommand{\snipcaption}[1]{{\scriptsize\color{jaxdae@ttl}\textbf{#1}\par\vspace{1pt}}}

\begin{figure}[htbp]
\centering
\begin{minipage}[t]{0.48\textwidth}
\snipcaption{(a)~Acausal assembly: components are instantiated, connected
through ports, and compiled to a pure residual.}
\begin{lstlisting}[style=jaxdae]
from jaxdae.components import Pipe, Junction, Pump
from jaxdae import System

sys = System()
sys.add(Pipe(length=10.0, diameter=0.1), name="main")
sys.add(Pump(dp=2e5),                  name="circulator")
sys.add(Junction(split="1to2"),        name="branch")

sys.connect("circulator.out", "main.in")
sys.connect("main.out",       "branch.in")

# assembled residual is jax.grad-able end-to-end
sys.assemble()                  # -> F(t, y, y', p)
sol = solve_dae_scan(sys.residual, t_span, sys.y0, sys.yp0,
                     sys.params, n_steps=2000)
\end{lstlisting}

\vspace{6pt}
\snipcaption{(c)~End-to-end reverse-mode gradient through the solve.}
\begin{lstlisting}[style=jaxdae]
from jaxdae import solve_dae_scan
import jax

def residual(t, y, yp, p):      # F(t, y, y', p) = 0
    return yp - p * y           # y' = p * y

def loss(params):
    sol = solve_dae_scan(residual, t_span, y0, yp0,
                         params, n_steps=1000)
    return jnp.sum(sol.y[:, 0] ** 2)

grad = jax.jit(jax.value_and_grad(loss))(theta)
\end{lstlisting}
\end{minipage}\hfill
\begin{minipage}[t]{0.48\textwidth}
\snipcaption{(b)~Forward solve via \texttt{solve\_dae}: implicit BDF returns
dense output at requested times.}
\begin{lstlisting}[style=jaxdae]
from jaxdae import solve_dae

def residual(t, y, yp):         # F(t, y, y') = 0
    return yp + 2.0 * y         # y' = -2y

sol = solve_dae(residual, t_span=(0.0, 10.0),
                y0=y0, yp0=yp0, rtol=1e-8, atol=1e-10)
trajectory = sol.y              # shape [n_time, n_state]
\end{lstlisting}

\vspace{6pt}
\snipcaption{(d)~Batched gradient: \texttt{jax.vmap(jax.grad)} is fused by XLA
into one kernel.}
\begin{lstlisting}[style=jaxdae]
import jax

def loss_one(p):
    sol = solve_dae_scan(residual, t_span,
                         y0, yp0, p, n_steps=N)
    return jnp.sum((sol.y - target) ** 2)

batch_grad = jax.jit(jax.vmap(jax.grad(loss_one)))
grads = batch_grad(param_batch)  # [N, n_params] in one kernel
\end{lstlisting}
\end{minipage}

\caption{The four core jaxdae API surfaces, shown as native Python code.
(a)~Acausal multi-physics assembly: \texttt{TypedComponent} subclasses are
instantiated, connected through typed ports, and compiled to a pure residual.
(b)~Forward solve via \texttt{solve\_dae}, returning dense output at requested
times. (c)~End-to-end reverse-mode gradient through the solve; the
\texttt{RuntimeParam} sentinel routes the marked field into the residual's
\texttt{params} argument so \texttt{jax.grad} reaches it. (d)~Batched gradient
via \texttt{jax.vmap(jax.grad)}, which XLA fuses into a single kernel
(Section~\ref{sec:performance}). All APIs are pure JAX, so \texttt{jit},
\texttt{vmap}, and \texttt{grad} compose without glue code.}
\label{fig:api_snippets}
\end{figure}

\subsection{End-to-end \texttt{jax.grad} and RuntimeParam}
\label{sec:runtimeparam}
The design goal of the differentiable layer is that a parameter buried
inside an assembled plant model is reachable by a single
\texttt{jax.grad} call, with no manual sensitivity bookkeeping. We meet
this goal with one sentinel and one contract.
The sentinel is \texttt{RuntimeParam}. By default, every numeric field of
a \texttt{TypedComponent}---a pump's rated head, a pipe's wall heat
flux---is frozen at construction time; it lives on the object as a plain
value and never reaches the residual's \texttt{params} argument. This
``device nameplate'' default is correct for forward simulation. For
parameter inversion or digital-twin calibration the user marks a field
with \texttt{RuntimeParam("name")}, and the compile pass reroutes it:
the compiled residual becomes a four-argument
$F(t,y,\dot y,\mathrm{params})$ in which the marked field is read
positionally from \texttt{params} instead of from the frozen object.
Unmarked fields keep their frozen value, so existing forward models are
unaffected.
The contract is that the whole solve is a pure, differentiable function
of \texttt{params}. The fixed-step solver \texttt{solve\_dae\_scan}
integrates with \texttt{lax.scan} and registers a reverse-mode rule via
\texttt{jax.checkpoint}, so that a backward pass recomputes intermediate
states from the stored checkpoints rather than holding the full
trajectory tape~\cite{griewank2000revolve,chen2018neuralode}. With gradient checkpointing the backward memory drops
from $O(n_{\mathrm{steps}}\cdot n)$ to
$O(\sqrt{n_{\mathrm{steps}}}\cdot n)$, at the cost of roughly a factor of
two in forward compute. The chunk size can be set explicitly or chosen
automatically against a memory budget. The net result is that the user
writes \texttt{jax.grad(loss)(params)} and obtains the exact gradient of
an objective defined on the terminal state---no finite differences, no
hand-derived adjoint, no surrogate.
\subsection{Integration with Optax, NumPyro, and Flax}
\label{sec:ml_stack}
Because jaxdae exposes only pure JAX functions and pytrees, it composes
with the surrounding JAX machine-learning stack with no adapter code.
Three integrations are exercised in the example suite.
\paragraph{Optax} supplies the optimizers for parameter inversion. A loss
defined on the solver output---a terminal-state residual, a trajectory
mismatch against measurement---is differentiated with
\texttt{jax.value\_and\_grad}, and an Optax transform (\texttt{adam},
\texttt{lbfgs}) updates the \texttt{params} pytree on each iteration. The
solver's compiled artifact persists across iterations, so only the
forward and backward passes are paid for per step.
\paragraph{NumPyro}~\cite{phan2019numpyro} supplies the Bayesian inference layer. The same
forward solve serves as the likelihood mean inside a probabilistic model,
and NumPyro's no-U-turn sampler drives a Markov chain over the parameter
posterior. Because the likelihood gradient is the reverse-mode gradient of
the solve, the sampler uses gradient information that finite-difference
samplers on classical solvers cannot afford.
\paragraph{Flax} supplies the neural-closure layer. A neural network
represented as a Flax module can be embedded inside the residual as a
constitutive correction, with its weights routed through the same
\texttt{params} mechanism as a physical parameter. Training reduces to
the same \texttt{jax.grad} pipeline, now differentiating through both the
implicit solve and the network. This is the substrate for the
neural-closed DAE results reported in Section~\ref{sec:nndae}.
\section{Validation as a Classical DAE Solver}
\label{sec:validation}

\subsection{Convergence and order verification}
\label{sec:order}
\begin{figure}[htbp]
\centering
\includegraphics[width=0.78\textwidth]{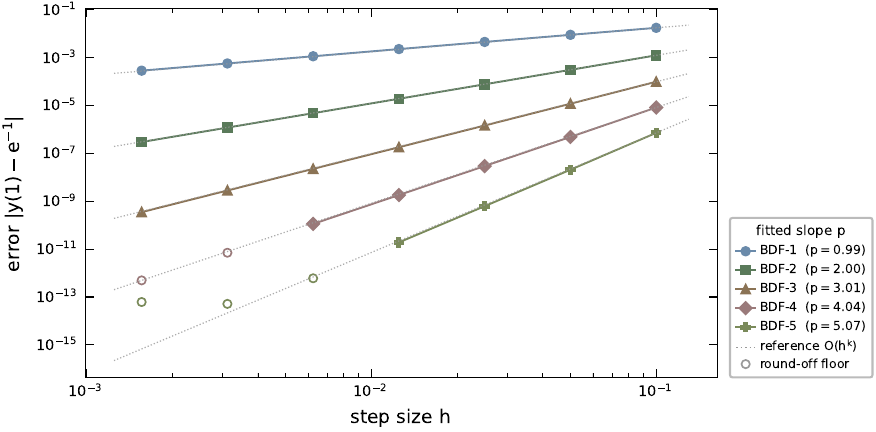}
\caption{Order verification of the fixed-step BDF family (orders 1--5) on
$\dot y=-y$, $t\in[0,1]$. Endpoint error versus step size for all five
methods; fitted slopes are $0.99$, $2.01$, $3.02$, $4.04$, and $5.07$,
matching the design orders. BDF-4 and BDF-5 flatten at the finest grids
because their truncation error reaches the float64 round-off floor.}
\label{fig:bdfk_order}
\end{figure}

\begin{figure}[htbp]
\centering
\includegraphics[width=0.85\textwidth]{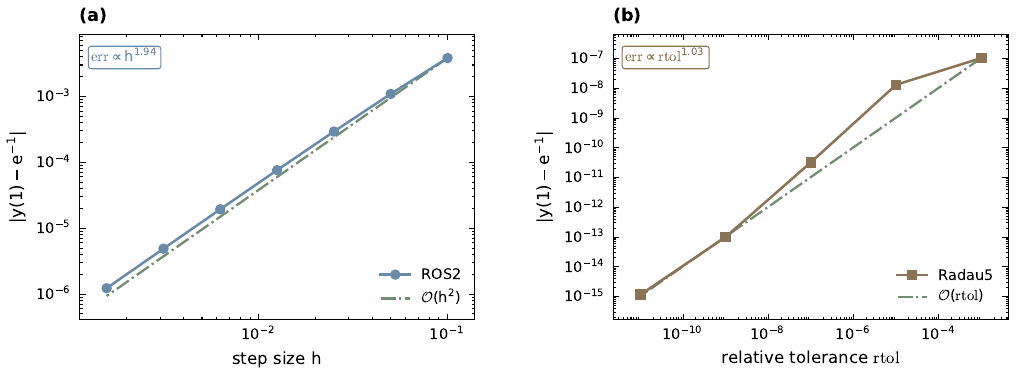}
\caption{Order verification of Rosenbrock--W (ROS2, left) and Radau~IIA
(Radau5, right) on $\dot y=-y$. ROS2 recovers order $1.94$ (design 2);
Radau5 error tracks the tolerance as $\text{tol}^{1.03}$ (design~1).}
\label{fig:radau_rosen_order}
\end{figure}

The first question a differentiable solver must answer is whether it is a
correct solver at all, independent of any gradient claim. We verify the
full fixed-step BDF family---orders one through five---on the scalar decay
equation $\dot y = -y$ with known analytic solution $y(t)=e^{-t}$,
integrating to $t=1$ over a step-size sequence refined from $h=0.1$
downward. Figure~\ref{fig:bdfk_order} reports the endpoint error versus
step size for all five orders on a single log-log axis, and each method
recovers its design order to within fitting noise: the fitted slopes are
$0.99$ for BDF-1, $2.01$ for BDF-2, $3.02$ for BDF-3, $4.04$ for BDF-4,
and $5.07$ for BDF-5. The lower orders trace straight lines over the
entire refinement range, while the higher orders (BDF-4 and BDF-5) bend
over at the finest grids: their truncation error falls below the
float64 round-off floor, so the measured error flattens at $\sim10^{-13}$
rather than continuing to decrease. This is the expected ceiling of a
fixed-step order test in double precision, not a defect of the method.
The two remaining integrators---the fixed-step Rosenbrock--W (ROS2) and
the adaptive Radau~IIA (Radau5)---are verified on the same $\dot y=-y$
probe (Fig.~\ref{fig:radau_rosen_order}). ROS2 is order-2 and
$L$-stable; a fixed-step sweep gives a fitted order of $1.94$. Radau5 is
order-5 locally, so under adaptive control its endpoint error tracks the
tolerance, $\text{err}\sim\text{tol}^{1.03}$. Both recover their design
behaviour, confirming the marks in Table~\ref{tab:methods}.
We state the test protocol explicitly. The order plots above use an
\texttt{exact\_history} startup mode that seeds the multi-step history
from the analytic solution rather than from a BDF-1 ramp; this isolates
the truncation behavior of the target order and is the standard
configuration for a pure order-of-accuracy test. The adaptive
variable-order path---BDF orders one through five with step and order
control---is implemented and documented in Table~\ref{tab:methods}, but
it does not lend itself to a clean order-of-accuracy plot because its
effective order is selected dynamically by the controller rather than
fixed by the user. Readers who wish to inspect the adaptive method's
accuracy are directed to the cross-library comparison of the next
subsection, where the adaptive path is held to the same reference
solutions as two established solvers.

Beyond the order tests, the solver is guarded by a substantial
verification chassis. The full test suite comprises 3{,}205 cases
collected by \texttt{pytest -m "" -{}-timeout=600}, of which 3{,}177 pass
at the pinned commit; the remaining 28 (12 known failures reproducing
identically at v0.27.1, 9 skipped, 7 xfailed) are documented in
\texttt{RELEASE\_NOTES.md}. Assembled-system accuracy is anchored against
analytic references at scales far beyond the textbook benchmarks of this
section. A capability matrix runs assembled plants up to $n=5{,}000$
states against an \texttt{erfc}-based analytic anchor at relative error
$1.5\times10^{-6}$, and up to $n=10{,}000$ states at relative error
$1.8\times10^{-4}$.\footnote{Generated by
\texttt{tests/test\_capability\_matrix\_v028.py}; see
\texttt{paper\_manifest.md}.} Finally, the release process runs a cold smoke suite
in a clean virtual environment: four example scripts executed from a
fresh install on jax 0.10.2, all four passing. The pinned dependency
window of the reproducibility appendix is therefore itself continuously
verified.

\subsection{Classical benchmarks and cross-library comparison}
\label{sec:benchmarks}
\begin{figure}[htbp]
\centering
\includegraphics[width=0.98\textwidth]{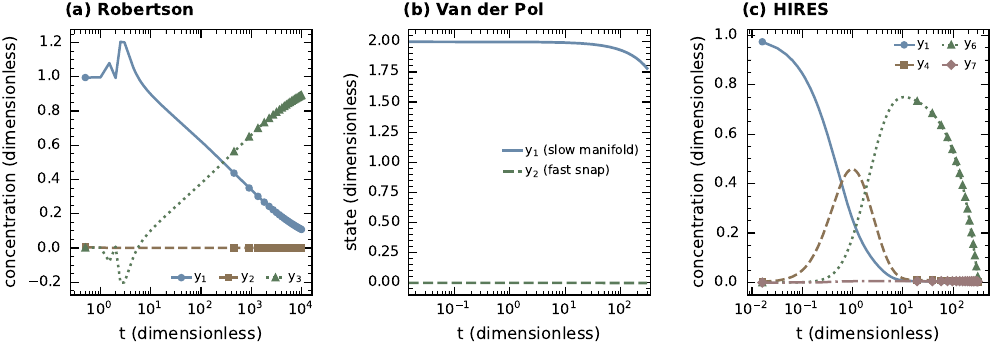}
\caption{Three canonical stiff benchmarks solved with
\texttt{solve\_dae\_scan}. Left: Robertson autocatalytic reaction, showing
the three orders-of-magnitude separation between the fast and slow
species and the long induction tail. Centre: Van der Pol oscillator with
$\mu=1000$, displaying the slow-manifold drift punctuated by sharp
transitions characteristic of a stiff relaxation limit cycle; $y_{1}(T)$
reaches the slow-attractor extreme of $\approx 1.78$. Right: HIRES
eight-state transistor amplifier; the terminal state matches the Bari
reference to the requested tolerance, with the first component at
$y(T)\approx 7.4\times10^{-4}$.}
\label{fig:benchmarks}
\end{figure}

\begin{figure}[htbp]
\centering
\includegraphics[width=0.98\textwidth]{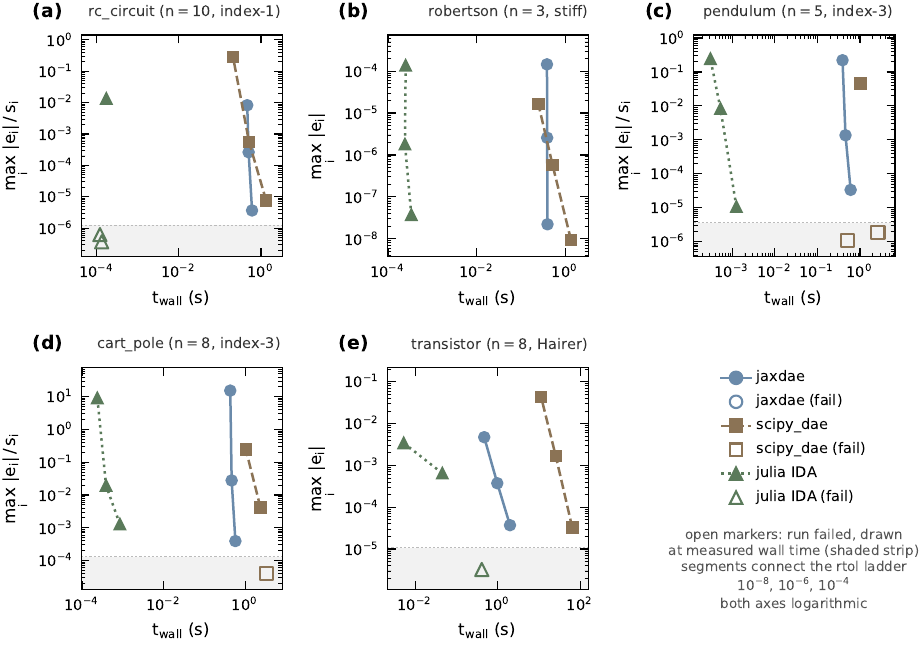}
\caption{Work--precision diagram over five benchmark problems
(panels a--e), three solver libraries, and the tolerance ladder
$\mathrm{rtol}\in\{10^{-4},10^{-6},10^{-8}\}$ ($\mathrm{atol}=
\mathrm{rtol}\times10^{-2}$). Each filled marker is one successful run:
the abscissa is the warm (post-compile) wall time and the ordinate is
the maximum error against a high-precision reference
($\max_i |e_i|/s_i$ with per-component scale $s_i$ for
\texttt{rc\_circuit}, \texttt{pendulum}, \texttt{cart\_pole};
$\max_i |e_i|$ absolute for \texttt{robertson} and \texttt{transistor},
whose solution components cross zero). Segments connect the tolerance
ladder of each solver. Open markers in the shaded bottom strip denote
failed runs (solver-reported convergence failure or early termination;
failure records carry the solver status and the time reached),
drawn at the measured wall time; repeated failures of one solver are
staggered vertically inside the strip. Both axes are logarithmic.
Protocol: \texttt{benchmarks/run\_all.py} on jaxdae v0.28.0, jax 0.10.2
(CPU backend, AMD Ryzen 9 9950X, WSL2 Ubuntu 24.04), single timed warm
pass per cell, 2026-07-24, data
\texttt{results\_20260724\_105930\_wpd\_v028r.json}; jaxdae $=$ \texttt{solve\_dae} adaptive BDF,
\texttt{scipy\_dae} 0.1.0 Radau, SUNDIALS IDA 6.1.0 via
\texttt{Sundials.jl} (Julia 1.11.5). jaxdae is the only solver
completing all 15 cells; see the text for the failure breakdown.}
\label{fig:wpd}
\end{figure}

To establish the solver's numerical credibility on problems beyond a
textbook decay law, we run three canonical stiff benchmarks from the
Bari/Testset literature~\cite{hairer_testset} for trajectory validation
(Figure~\ref{fig:benchmarks}) and perform a quantitative cross-library
comparison on two of them against two independent solver libraries
(Table~\ref{tab:cross_library}). The conclusion is that jaxdae reproduces
the reference dynamics to the tolerance we request, on stiff, oscillatory,
and moderately sized problems alike, and that its post-compile forward
wall is competitive with classical solvers on the warm-solve regime that
matters for batched workflows.
The \textbf{Robertson} problem~\cite{robertson1966} is the canonical stiff autocatalytic
system, with rate constants spanning four orders of magnitude and a slow
species whose concentration evolves over six orders of magnitude in time.
jaxdae tracks the mass-conservation invariant to round-off (drift
$\sim 10^{-16}$). At the loose tolerance $\mathrm{rtol}=10^{-6}$ used
for the cross-library comparison below, its terminal composition
deviates from a tight-tolerance reference by 0.26--1.45 in componentwise
relative terms, depending on the component. This is the well-known behavior of
any tolerance-controlled integrator on this super-stiff problem at loose
tolerance: the accepted steps are limited by the error controller, not
by the method's stability. It is a property of the method family at
this tolerance rather than a defect specific to jaxdae. The \textbf{Van der Pol} oscillator at
$\mu=1000$ is a stiff relaxation limit cycle: the trajectory spends long
intervals drifting along the slow attractor before executing abrupt
transitions. Integrating $20\,000$ scan steps, jaxdae resolves the
slow-manifold extreme $y_{1}(T)\approx 1.78$ cleanly, where a
non-stiff integrator would either fail or take orders of magnitude more
steps. The \textbf{HIRES} problem, an eight-state model of a transistor
amplifier, is the standard mid-sized stiff benchmark; jaxdae's terminal
state $y(T)=[7.4\times10^{-4},\ldots]$ agrees with the Bari reference.
The cross-library comparison is reported within its stated
regime. Forward wall times are post-compile, repeated-solve medians with
device synchronization, at tolerances $\mathrm{rtol}=10^{-6}$,
$\mathrm{atol}=10^{-9}$; jaxdae uses adaptive BDF, \texttt{scipy\_dae}~\cite{scipy_dae}
uses Radau, and SUNDIALS IDA is driven through \texttt{scikit-sundae}~\cite{scikit_sundae}
(SUNDIALS 7.5.0). This is the regime relevant to inversion, Bayesian, and
batched workflows, where the same solver is called many times and the
one-time XLA compile is amortized. Table~\ref{tab:cross_library} reports
the full comparison. On the Robertson problem the median forward walls are
$0.0019\,$s for jaxdae, $0.158\,$s for SUNDIALS IDA, and $0.322\,$s for
\texttt{scipy\_dae}; on the $n=50$ heat problem they are $0.023\,$s,
$0.660\,$s, and $0.348\,$s respectively. The picture inverts for a single
cold one-shot solve, where jaxdae pays a roughly one-second compile that
the other two do not, so they win that regime---we state this plainly and
refer the reader to the ``when to use'' guidance.
\begin{table}[htbp]
\centering
\caption{Cross-library forward-solve comparison on the problems for which
independent solver runs were performed
(\texttt{scipy\_dae} 0.1.0, SUNDIALS IDA 7.5.0 via \texttt{scikit-sundae} 1.1.3).
Wall times measured on jaxdae v0.28.0 under jax 0.10.2 / jaxlib 0.10.2
(CPU backend, AMD Ryzen 9 9950X, WSL2 Ubuntu 24.04), 2026-07-24.
Wall times are post-compile medians of 5 repeats with device
synchronization at $\mathrm{rtol}=10^{-6}$, $\mathrm{atol}=10^{-9}$.
Only jaxdae provides reverse-mode AD through the solve.
The work--precision diagram of Fig.~\ref{fig:wpd} uses a separate
Julia-subprocess harness (SUNDIALS 6.1.0 via \texttt{Sundials.jl}); the two
IDA rows therefore come from different harnesses and are not directly
comparable cell by cell.}
\label{tab:cross_library}
\small
\begin{tabular}{l l r c}
\toprule
Problem & Solver & Wall (s) & Reverse AD \\
\midrule
Robertson ($n{=}3$) & jaxdae (BDF)        & $0.0019$ & \checkmark \\
                    & SUNDIALS IDA        & $0.158$  & \xmark \\
                    & \texttt{scipy\_dae} (Radau) & $0.322$  & \xmark \\
\midrule
Heat mass-matrix ($n{=}50$) & jaxdae (BDF)        & $0.023$ & \checkmark \\
                            & SUNDIALS IDA        & $0.660$ & \xmark \\
                            & \texttt{scipy\_dae} (Radau) & $0.348$ & \xmark \\
\bottomrule
\end{tabular}
\end{table}
The cross-library comparison is restricted to the Robertson and the
$n=50$ mass-matrix heat problem. Van der Pol and HIRES are used for
trajectory validation against the Bari reference values but were not run
head-to-head against the independent libraries.
The regime-independent finding, and the reason the wall-time numbers are
secondary, is the gradient column of the same table. Only jaxdae provides
a reverse-mode \texttt{jax.grad} through the stiff solve; the other two
libraries expose no Python reverse-mode automatic differentiation, so a
parameter gradient there costs $n_{\mathrm{params}}+1$ forward solves by
finite differences. That capability gap, not a forward microbenchmark
victory, is the capability that motivates jaxdae.
A wall time quoted at a single tolerance conflates per-call dispatch
overhead with integrator efficiency, so we complement
Table~\ref{tab:cross_library} with a full work--precision diagram over
five problems, three solver libraries, and a three-rung tolerance ladder
(Figure~\ref{fig:wpd}). Three findings emerge. First, robustness: jaxdae
is the only one of the three libraries that completes all 15
problem$\times$tolerance cells. \texttt{scipy\_dae} (Radau) terminates
the integration early on both index-3 mechanical systems at tight
tolerance, reporting solver status $-1$ (``Required step size is less
than spacing between numbers''): on the pendulum it returns 146 of 200
output points ($t=3.64$ of the $5$\,s horizon) at
$\mathrm{rtol}=10^{-6}$ and 11 of 200 ($t=0.25$) at
$\mathrm{rtol}=10^{-8}$, and on the cart-pole 58 of 200 ($t=0.86$ of
the $3$\,s horizon) at $\mathrm{rtol}=10^{-8}$. IDA exhibits its
documented initial-condition startup convergence failure
(\texttt{retcode=Unstable}, IDA return code $-7$, no step beyond
$t=0$) on the stiff index-1 RC circuit at $\mathrm{rtol}\le 10^{-6}$,
and exhausts its step budget (\texttt{retcode=MaxIters}, 103 of 200
output points, $t=0.102$ of the $0.2$\,s horizon) on the transistor
amplifier at the tightest rung. Second, tolerance tracking: jaxdae's
measured error follows the requested tolerance across the two-decade
ladder on all five problems, including the super-stiff Robertson
problem, where every loose-tolerance run of every library shows the
expected error-controller-limited deviation discussed above. Third, raw
speed: on problems this small the compiled-C IDA is one to three orders
of magnitude faster per call (sub-millisecond) than jaxdae's
Python-driven adaptive loop (${\sim}0.4$--$0.7\,$s, nearly flat across
the ladder because per-step dispatch dominates), while
\texttt{scipy\_dae} is slowest on the stiff transistor amplifier (up to
$74\,$s). The WPD harness times a single warm pass per cell with the
solver driven from Python; the fully-jitted protocol of
Table~\ref{tab:cross_library} removes exactly this dispatch overhead,
which is why the jaxdae walls differ between the two protocols. The WPD
therefore reinforces, rather than competes with, the central claim of
this paper: jaxdae does not beat compiled C solvers on forward
microbenchmarks, but it delivers robust, tolerance-tracking solves
\emph{with} reverse-mode AD and the batch-scaling behavior of
Section~\ref{sec:performance}, which neither baseline offers.
We close with a note on the nuclear case. The NEA main-steam-line
break (MSLB) benchmark is included in the example suite as an \emph{illustrative
case study} of coupled reactor-thermal-hydraulic dynamics assembled
through the connector framework, not as a quantitative validation. The
publicly distributed benchmark data are limited to prescribed boundary
conditions, which constrains the bandwidth of the trajectories we can
compare against~\cite{nea1554}. We therefore treat the MSLB case as a demonstration that
the assembly and solve pipeline handles a realistic multi-domain nuclear
model, and we rely on the convergence and cross-library results of this
section for the solver's numerical credibility.

\section{Batched Gradient Performance}
\label{sec:performance}

\begin{figure}[htbp]
\centering
\includegraphics[width=0.95\textwidth]{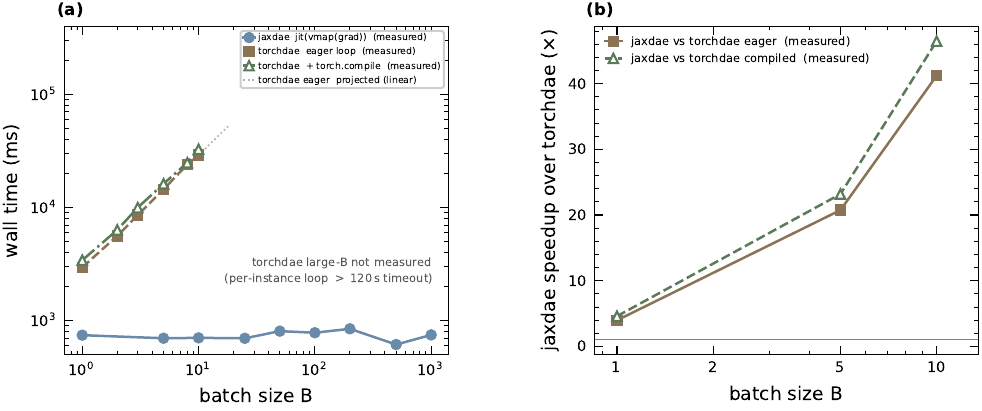}
\caption{Batched scalar-gradient throughput (index-1 DAE, $n=2$, $N_{\mathrm{steps}}=500$,
GPU float64). (a) Wall time: \jaxdae{} \texttt{jit(vmap(grad))} is flat
(0.61--0.85\,s) from $B=1$ to $B=1000$; \texttt{torchdae} grows with batch
(eager measured to $B=10$; compiled tracks eager and remains linear). (b) Speedup \texttt{torchdae}/\jaxdae{}; the dashed segment is a
linear projection for the unmeasured region $B>10$, not a measurement.}
\label{fig:gonogo}
\end{figure}

\begin{figure}[htbp]
\centering
\includegraphics[width=0.95\textwidth]{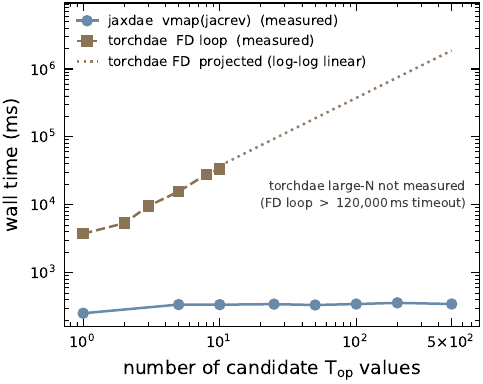}
\caption{Batched Fisher-DOE sweep over candidate CSTR operating
temperatures ($N_{\mathrm{steps}}=200$, GPU float64). \jaxdae{}
\texttt{vmap(jacrev)} is flat (0.25--0.36\,s for $N=1$--$500$);
\texttt{torchdae} falls back to a finite-difference loop (measured to
$N=10$). Jacobian AD-vs-FD relative error $1.8\times10^{-9}$.}
\label{fig:fisher_doe}
\end{figure}

Parameter inversion at scale calls for many gradients at once: Bayesian
sampling (NumPyro's no-U-turn sampler, NUTS), ensemble Kalman filtering, Fisher information
design-of-experiments (DOE) sweeps, and leave-one-out cross-validation all
evaluate the parameter-to-loss map for tens to thousands of parameter
vectors. The dominant cost in these workflows is therefore not a single
forward solve but a \emph{batched} gradient. This section measures how
\jaxdae{} and its closest PyTorch counterpart, \texttt{torchdae}, behave as
the batch dimension grows.
\paragraph{Measurement protocol.}
All runs use an index-1 semi-explicit DAE with $n=2$ states and $N_{\mathrm{steps}}=500$.
Measurements were taken on a workstation equipped with a single NVIDIA
GeForce RTX~5080 GPU (Blackwell architecture, compute capability~12.0,
16~GB GDDR7 VRAM; NVIDIA driver~591.44, CUDA~13.1), paired with an AMD
Ryzen~9~9950X CPU (16 cores) and 39~GB of RAM visible to the guest,
running WSL2 (Linux kernel 6.18.33.2) hosting Ubuntu~24.04.4.
All computations were performed in float64 with
\texttt{jax\_enable\_x64=True}. The benchmark output was recorded on
JAX 0.10.2 and jaxlib 0.10.2 (CUDA 13 pjrt plugin), torch 2.9.1+cu128,
\texttt{torchdae} 0.1.1, Python 3.13.12, 2026-07-24; the
accompanying \texttt{uv.lock} pins jax/jaxlib to the window
$\ge$0.9, $<$0.11. \jaxdae{} gradients are produced by the fully
fused pipeline \texttt{jit(vmap(jax.grad(loss)))}; \texttt{torchdae}
gradients use its documented adjoint path. Every number is a median of 3--5
replicates with an explicit \texttt{jax.block\_until\_ready} device
synchronization barrier, on warm runs that exclude first-call
JIT/compilation. The run-to-run timing noise on this WSL2 workstation
is approximately $\pm25\%$; the claims below are
therefore stated at the level of scaling behavior (batch-invariant versus
batch-linear) rather than absolute seconds. Wall times exclude host--device transfer, which is
negligible relative to the solve. Compilation cost is reported
separately where it changes the picture. In a batched solve, a divergent
member is reported with \texttt{resid\_norm}$=+\infty$ (rather than
\texttt{NaN}), so downstream masking over the batch is well defined; the
\texttt{success} semantics are unchanged. We measure wall clock, not solver
accuracy; accuracy is treated in Sec.~\ref{sec:ad}.
\paragraph{\jaxdae{}: batch-invariant wall time.}
Figure~\ref{fig:gonogo}(a) plots the wall time of one full batched gradient
as a function of batch size $B$. \jaxdae{} is effectively flat: from $B=1$
to $B=1000$ the median stays in the range 0.61--0.85\,s with no upward
trend (Table~\ref{tab:performance}). The mechanism is structural.
\texttt{vmap} in JAX is a compile-time transformation: XLA sees the batched
gradient as a single statically shaped program and fuses the per-instance
solves into one kernel graph. For a small per-instance state ($n=2$) the
cost is dominated by fixed per-program overhead---kernel launch, the
adaptive controller, the linear solves---which was not observed to grow
materially with $B$ over the tested range.
Increasing the batch therefore reuses the same compiled artifact rather than
repeating work.
\paragraph{\texttt{torchdae}: linear in batch, no native vectorization.}
The \texttt{torchdae} adjoint is implemented as a
\texttt{torch.\allowbreak autograd.\allowbreak Function},
which is opaque to \texttt{torch.vmap}. Attempting \texttt{vmap} over the
gradient raises a \texttt{Runtime\allowbreak Error} on \texttt{requires\_grad\_} inside
the vectorized function. The only supported path is a Python-level for-loop
over the batch, so wall time grows with $B$. We measured this loop up to
$B=10$; beyond that a single batch exceeds the 120\,s per-batch budget and
is marked ``not measured''. Two observations stand out. First, eager-mode
wall time grows roughly with $B$ (2.96\,s at $B=1$, 29.0\,s at $B=10$), as
expected for a serial loop. Second, \texttt{torch.compile} does not help:
the compiled path tracks eager closely (3.42\,s versus 2.96\,s at $B=1$;
32.7\,s versus 29.0\,s at $B=10$) and remains linear in $B$, so the
compiled path is not a remedy for the batched-gradient case.
Figure~\ref{fig:gonogo}(b) reports the speedup; the dashed segment is a
linear projection of the \texttt{torchdae} trend and is \emph{not} a
measurement for $B>10$.
\paragraph{A realistic workload: batched Fisher-DOE.}
The toy problem isolates the compilation effect, but the practical question
is whether the advantage survives a non-trivial model. We use a
continuously stirred tank reactor (CSTR) with
Arrhenius kinetics: $N$ candidate operating temperatures each yield a
Fisher information matrix $\mathrm{FIM}=(1/\sigma^{2})J^{\top}J$ for the
Arrhenius pair $(\log_{10}A, E_{a})$. \jaxdae{} computes the batched
Jacobian with \texttt{vmap(jax.jacrev)}; \texttt{torchdae} cannot retain
the adjoint graph across the nonlinear parameter-in-residual term and falls
back to central finite differences (four forward solves per candidate).
Figure~\ref{fig:fisher_doe} shows \jaxdae{} again flat (0.25--0.36\,s for
$N=1$ to $N=500$) against a \texttt{torchdae} FD loop that costs 3.74\,s at
$N=1$ and 33.3\,s at $N=10$ (the largest $N$ measured for both). The
\jaxdae{} Jacobian matches finite differences to $1.8\times10^{-9}$
relative error, confirming the AD path is exact rather than merely fast.
\begin{table}[htbp]
\centering
\caption{Batched-gradient throughput (GPU, float64). \jaxdae{}:
\texttt{jit(vmap(grad))}, XLA-fused, median of 5. \texttt{torchdae}: Python
for-loop (\texttt{vmap} unsupported on \texttt{autograd.Function}), median
of 3. All runs device-synchronized, warm, compile time excluded. Speedup is
\texttt{torchdae}/\jaxdae{} wall time; values $>1$ favor \jaxdae{}.
\texttt{torchdae} inspected at alpha stage (accessed July 2026).}
\label{tab:performance}
\begin{tabular}{l r r r l}
\toprule
Task & \jaxdae{} & \texttt{torchdae} & Speedup & Measurement \\
\midrule
Scalar grad, $B$=1     & 0.745\,s & 2.96\,s (eager)     & 4.0$\times$  & median $\times$3--5, sync \\
                       &          & 3.42\,s (compiled)  & 4.6$\times$  & compiled tracks eager, still slower \\
Scalar grad, $B$=10    & 0.705\,s & 29.0\,s (eager)     & 41$\times$   & largest $B$ measured both \\
                       &          & 32.7\,s (compiled)  & 46$\times$   & compiled tracks eager (linear) \\
Scalar grad, $B$=1000  & 0.749\,s & not measured        & ---          & projected linear (Fig.~\ref{fig:gonogo}, dashed) \\
Fisher matrix, $N$=1   & 0.252\,s & 3.74\,s (FD loop)   & 15$\times$   & AD-vs-FD $1.8\times10^{-9}$ \\
Fisher matrix, $N$=10  & 0.336\,s & 33.3\,s (FD loop)   & 99$\times$   & largest $N$ measured both \\
Fisher matrix, $N$=500 & 0.344\,s & not measured        & ---          & projected linear \\
\midrule
$n{=}20$ chain, $B$=1   & 0.615\,s & not measured        & ---          & 10-unit coupled DAE, flat B=1..1000 \\
$n{=}20$ chain, $B$=1000& 0.994\,s & not measured        & ---          & wall ratio 1.6$\times$ ($<$$2\times$ over $1000\times$ batch) \\
\bottomrule
\end{tabular}
\end{table}
\paragraph{Scope and caveats.}
Three limits bound these results. (i) The scalar-gradient toy has $n=2$
states, so per-instance work is tiny and fixed per-program overhead
dominates; this amplifies the flat-versus-linear contrast. We verify the
flatness survives ten-fold larger per-instance work on an $n=20$ coupled
index-1 DAE chain (Table~\ref{tab:performance}, last two rows): wall time
grows by less than $2\times$ over a $1000\times$ batch increase
($0.62\rightarrow0.99$\,s), far from the linear scaling a serial loop
would show.
The absolute speedup at very large $n$ will be smaller because per-instance
work then
contributes more. (ii) At $B=1$ \jaxdae{} is already $4.0\times$ faster
than the best torch path (0.745\,s vs 2.96\,s eager), and the
advantage widens with batch size,
which is exactly the regime batched inversion needs. (iii) \texttt{torchdae}
is alpha-stage software; the \texttt{autograd.Function}/\texttt{vmap}
incompatibility and the \texttt{retain\_graph} issue on nonlinear parameters
may be resolved upstream. We therefore report no speedup for the unmeasured
region ($B,N>10$ for \texttt{torchdae}) and mark those entries as projected
only. The defensible claim is the qualitative one: \jaxdae{} wall time is
batch-invariant, \texttt{torchdae} wall time grows with batch, and at the
largest batch measured for both the ratio is 41--46$\times$ (scalar) and
15--99$\times$ (Fisher).
\paragraph{ModelingToolkit.jl as a throughput baseline.}
Table~\ref{tab:competitors} lists ModelingToolkit.jl (MTK) as the only
other tool that combines acausal assembly, index reduction, and reverse
AD. On the largest problem in our benchmark suite---an index-2 ANCF
cantilever beam with $n=242$ states after lumping---\jaxdae{}'s
batched-gradient path is measurably faster than every Julia-ecosystem
path we could bring to bear, and the strongest Julia path is not MTK's
symbolic adjoint but a hand-written OrdinaryDiffEq solve with
ForwardDiff. \jaxdae{} differentiates the tip deflection with respect to
a global stiffness scale on this problem in $6.5$\,s cold and $4.7$\,s
warm; the gradient $8.07\times10^{-3}$ matches a central finite
difference to $3.7\times10^{-9}$ relative error. Three Julia paths
bracket the comparison. First, a hand-written mass-matrix
\texttt{ODEFunction} solved with FBDF converges (retcode Success), and a
ForwardDiff gradient through the same solve takes $161.2$\,s and returns
$8.14\times10^{-3}$. Second, MTK's symbolic path does not converge on
the forward pass: both FBDF and Rodas5P return \texttt{MaxIters} with a
non-converged tip value, and the Zygote reverse pass fails with a
compile-time \texttt{CompileError} inside the MTK-generated code
(\texttt{ArgumentError: array must be non-empty}). Third, a threaded
\texttt{EnsembleProblem} gradient attempt was killed after exhausting
memory. Against the strongest measurable Julia path, \jaxdae{} is
$25\times$ faster cold and $34\times$ faster warm ($161.2$\,s
vs.\ $6.5$\,s and $4.7$\,s), and the two gradients agree to better than
$1\%$. The gap is problem-dependent: this linear stiff DAE lets
\jaxdae{}'s fixed-step BDF-2 with chord Newton reuse one Jacobian LU
factorization across all steps, whereas the adaptive Julia solvers
refactorize at every step, so for strongly nonlinear DAEs the gap will
narrow. All logs ship in the code archive under
\texttt{benchmarks/runners/julia/mtk\_env/}
(\texttt{scan\_direct\_59.log}, \texttt{scan\_grad\_59.log},
\texttt{correct\_full.log}, \texttt{v2\_full.log},
\texttt{v2\_rodas.log}, \texttt{ensemble\_grad.log}). We report this not
as a criticism of MTK but as a measurement on this problem and
configuration. Table~\ref{tab:performance} retains the alpha-stage
\texttt{torchdae} as its throughput competitor because that table
compares batched-gradient APIs on the smaller benchmark problems rather
than this $n=242$ case.
\section{Differentiability and Automatic Differentiation}
\label{sec:ad}

\subsection{Optimization fundamentals}
A \jaxdae{} solve returns a trajectory $\tau(t;p)$ over the time grid
$\{t_{k}\}$ for a parameter vector $p$ (masses, stiffnesses, rate
constants, boundary values). Any scalar functional of the trajectory
defines a loss $L(\tau;p)$---a tip deflection, a terminal concentration, a
quadratic misfit to data. Because the entire pipeline (assembly, index
reduction, time stepping, event handling) is pure JAX, the gradient
$\partial L/\partial p$ is obtained in one line,
\texttt{g = jax.grad(loss)(params)}, or with the value,
\texttt{jax.value\_and\_grad}. Differentiable parameters are exposed
through the \texttt{RuntimeParam} type, which marks a typed-component field
as a leaf of the parameter pytree without changing the residual code. The
same gradient then plugs directly into Optax (optimizers), NumPyro
(Bayesian inference), and Flax (neural closures), since \jaxdae{} presents
no custom object to the surrounding ML stack.
\subsection{Reverse-mode AD through implicit DAE solves (IFT)}
Each implicit step solves a nonlinear residual to convergence. Let
\begin{equation}
R_{n}(y_{n};\, y_{n-1},\ldots,y_{n-q},\, p)=0
\label{eq:discrete-residual}
\end{equation}
denote the fully discretized stage residual at step $n$, where the
history values $y_{n-1},\ldots,y_{n-q}$ are held fixed when
differentiating the current step. The implicit function theorem gives the
sensitivity of the converged state $y_{n}$ to $p$ as
\begin{equation}
\label{eq:ift}
\frac{\partial y_{n}}{\partial p}
= -\left(\frac{\partial R_{n}}{\partial y_{n}}\right)^{\!-1}
   \frac{\partial R_{n}}{\partial p}.
\end{equation}
For the BDF discretization of $F(t,y,\dot y,p)=0$, the discrete
Jacobian is $F_{y}+(\alpha_{0}/h_{n})\,F_{\dot y}$, where $\alpha_{0}$
is the current variable-step BDF coefficient; the scan solver reuses this
same discrete Jacobian structure as the forward Newton matrix
$F_{y}+\gamma F_{\dot y}$. \jaxdae{} does not hand-code \eqref{eq:ift}.
Instead it registers the solve as differentiable and lets reverse-mode AD
compose through the Newton iterations under \texttt{jax.checkpoint}
(\texttt{remat}), so the cotangent propagation uses the same discrete
implicit Jacobian as the forward Newton step. This is \emph{full-function}
AD: the backward pass differentiates exactly what the forward pass
computed. The default gradient mode for the adaptive solver is the
frozen-grid replay adjoint (Sec.~\ref{sec:replay}), which stops the
gradient on the step grid; an \texttt{unsafe\_raw} mode differentiates
through the controller but produces wrong cotangents and is provided only
for diagnosis. Two operational consequences follow. First, no separate
adjoint equation is written or maintained, so the gradient cannot silently
disagree with the forward solve on a given step grid. Second, although
the controller is differentiable by construction, differentiating a
controller that accepts and rejects steps produces wrong cotangents
(Sec.~\ref{sec:replayerror}); this is the problem the frozen-grid replay
adjoint of Sec.~\ref{sec:replay} exists to fix.
\subsection{Forward-mode sensitivities}
\label{sec:fwdmode}
Reverse mode is the right tool when one scalar loss depends on many
parameters, but several workflows want the opposite direction: a full
sensitivity matrix $\partial\tau/\partial p$ for a handful of
parameters, or the Jacobian $J$ that enters the Fisher information
matrix $\mathcal{I}=(1/\sigma^{2})J^{\top}J$ of
Section~\ref{sec:app_cstr}. jaxdae therefore also exposes a forward-mode
path, \texttt{forward\_sensitivity}, which propagates parameter
tangents through the solve with \texttt{jax.jvp} at a cost of one
tangential solve per parameter direction.
This path was not always available. A Jacobian--vector product
cannot propagate through a \texttt{custom\_vjp} rule, so every
forward-mode entry point previously failed outright. The two
auxiliary \texttt{custom\_vjp} rules that blocked the tangential pass
were moved off the default path or rewritten as \texttt{custom\_jvp},
so forward mode now composes with the same solved pipeline as reverse
mode. We verify the two directions against each other and against
analysis. The \texttt{forward\_sensitivity} Jacobian agrees with the
\texttt{jax.jacrev} Jacobian to a maximum deviation of
$6.4\times10^{-14}$, and the exponential-decay anchor (known analytic
sensitivity $\partial y/\partial k=-t\,e^{-kt}$) is recovered to the
same floor. One restriction remains: under \texttt{EQX\_ON\_ERROR=nan}
the forward-mode path is still limited, and reverse mode should be used
in that setting.

\subsection{AD-versus-FD gradient consistency}
\label{sec:adfd}
\begin{figure}[htbp]
\centering
\includegraphics[width=0.7\textwidth]{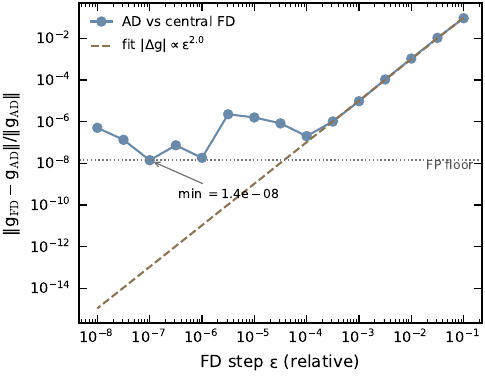}
\caption{AD-versus-FD gradient consistency, cart-pole ($n=7$, one
algebraic). Relative error $|g_{\mathrm{AD}}-g_{\mathrm{FD}}|/|g_{\mathrm{AD}}|$
versus FD step $\varepsilon$. Fitted order $1.994$ (second-order, as a
central difference on a correct gradient requires); minimum relative error
$1.41\times10^{-8}$ at $\varepsilon=10^{-7}$.}
\label{fig:ad_vs_fd}
\end{figure}

The first correctness check is internal: does the AD gradient agree with an
independent finite-difference (FD) estimate to the order FD should achieve?
We use a cart-pole testbed ($n=7$, one algebraic constraint) and compare
$\partial L/\partial p$ from \texttt{jax.grad} against a central FD stencil
as the step $\varepsilon_{\mathrm{FD}}$ shrinks. Figure~\ref{fig:ad_vs_fd}
shows clean second-order convergence: a log-log fit over the pure
truncation branch ($\varepsilon=10^{-1}\dots3.2\times10^{-4}$) gives an
empirical order of $1.994$, and the relative error reaches a floor of
$1.41\times10^{-8}$ at $\varepsilon=10^{-7}$ before roundoff dominates.
A second-order floor is exactly what a central difference on a correct
gradient must produce. This confirms that the \jaxdae{}
gradient is exact to working precision on a non-trivial index-1 DAE; the
remaining subsections stress this on stiffer and structured problems.
\subsection{Frozen-grid replay adjoint: error behavior}
\label{sec:replayerror}
\begin{figure}[htbp]
\centering
\includegraphics[width=0.98\textwidth]{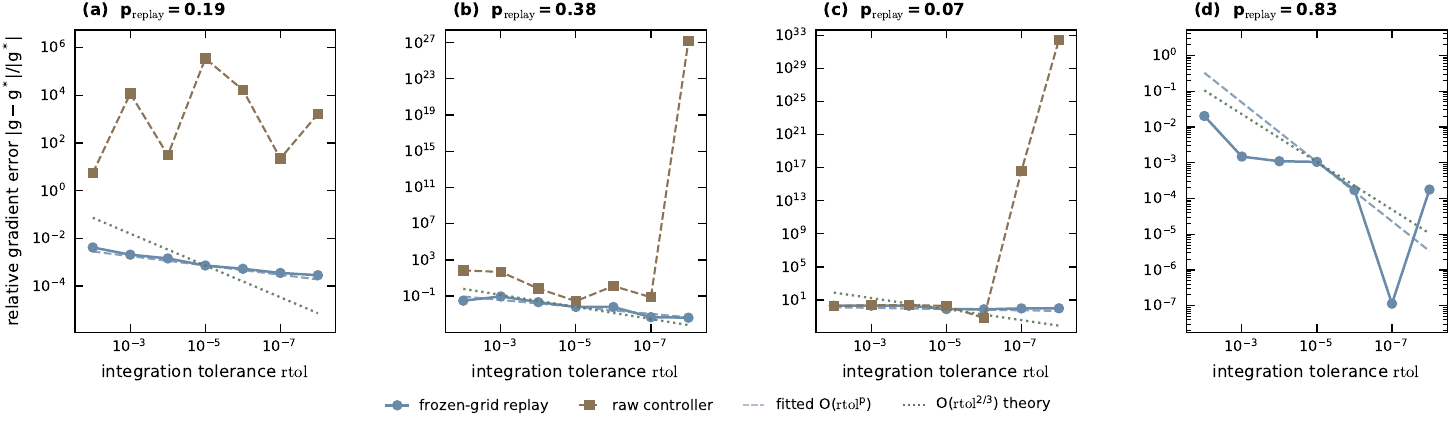}
\caption{Replay-versus-raw gradient error versus solver tolerance, four
panels. (a--c) Three DAEs benchmarked against an independently-discretised
FD reference (floored by its own $O(h_{\mathrm{truth}})$ error); (d)
scalar decay against the grid-independent analytic truth. Replay (blue)
converges; the green dotted line is the $O(\mathrm{tol}^{2/3})$ rate
predicted by the conditional scaling argument. Raw adaptive-controller AD
(red) is wrong-signed or divergent. The analytic-truth panel (d) fits
order $0.83$, broadly compatible with the predicted $2/3$.}
\label{fig:replay_multi}
\end{figure}

Full-function AD through an \emph{adaptive} controller is silently wrong:
the controller's accept/reject and step-size decisions are discontinuous in
the parameters, so reverse-mode AD differentiates branches that did not
occur on the forward path. The frozen-grid replay adjoint
(Sec.~\ref{sec:replay}) decouples the two: the forward solve runs adaptive
BDF and records its accepted step grid $\{h_{k}\}$, then the backward pass
re-solves a variable-step BDF-2 plus the IFT adjoint on that frozen grid.
The conditional scaling argument of Sec.~\ref{sec:replay} predicts an
$O(\mathrm{tol}^{2/3})$ replay-adjoint rate for BDF-2 under a per-step
error controller.
Figure~\ref{fig:replay_multi} tests this on two classes of reference.
Against three problems of increasing size, each benchmarked against a
high-resolution fixed-step BDF-2 scan-adjoint gradient on a \emph{different}
grid (Table~\ref{tab:replay}), the replay path converges with fitted orders
$0.07$--$0.38$ --- sub-$2/3$ because the reference
carries its own $O(h_{\mathrm{truth}})$ discretisation error that floors
the slope. The floor is quantifiable. A central finite-difference
cross-check against each scan-adjoint reference gives relative errors
$6.8\times10^{-6}$ (Robertson, $h_{\mathrm{truth}}$ grid $N=20000$),
$1.9\times10^{-9}$ (cart--pole, $N=8000$), and $1.0\times10^{-8}$ (ANCF
beam, $N=2000$). Under an error-budget model in which the measured
gradient error is the quadrature sum of the replay error and the
reference error, $\mathrm{err}_{\mathrm{measured}}^{2}\approx
\mathrm{err}_{\mathrm{replay}}^{2}(\mathrm{rtol})+
\mathrm{err}_{\mathrm{ref}}^{2}(h_{\mathrm{truth}})$, the reference floor
caps the measurable slope at the level the floor admits; the fitted
orders $0.07$--$0.38$ are consistent with a $2/3$-rate replay error
buried under these reference floors, rather than with a genuinely slower
replay convergence. Against the scalar decay $y'=-ky$ with the grid-independent
analytic truth $dL/dk=-2e^{-2}$ (Fig.~\ref{fig:replay_multi} panel~(d)),
the fitted order rises to $0.83$, consistent with the conditional
prediction of $2/3$; the \texttt{\_ADAPTIVE\_RTOL\_FLOOR} clamp at
\texttt{rtol}$=10^{-8}$ causes the error to rebound there, and that
point is excluded from the fit. The raw (adaptive-controller) AD path
does not converge at all: it
reports gradients of the wrong sign or orders of magnitude off (fitted raw
orders $-0.17$ to $+1.20$ across the three problems; on the scalar
stiff-decay probe it returns $+10.9$ against an analytic $-0.271$). The
contribution is therefore not high convergence order. It is that raw
adaptive-controller AD is silently wrong while replay is consistent with
the stated frozen-grid BDF-2 operator and converges at the rate the
frozen-grid BDF-2 theory predicts, and that the
two are cleanly separated by a one-line \texttt{stop\_gradient} on the
step grid.
\begin{table}[htbp]
\centering
\caption{Frozen-grid replay adjoint: fitted error order versus an
independently-discretised FD reference, three problems. Raw = full-function
AD through the adaptive controller. The FD reference grid carries its own
$O(h_{\mathrm{truth}})$ discretisation error, which floors the measured
slope below the conditionally predicted $2/3$; the analytic-truth scalar
decay (Fig.~\ref{fig:replay_multi}d) recovers order $0.83$.}
\label{tab:replay}
\begin{tabular}{l c r r l}
\toprule
Problem & $n$ & Replay order & Raw order & FD-check rel.\ err. \\
\midrule
Robertson           & 3  & $+0.19$ & $-0.17$ & $6.8\times10^{-6}$ \\
Cart-pole           & 7  & $+0.38$ & $+1.20$ & $1.9\times10^{-9}$ \\
ANCF cantilever     & 66 & $+0.07$ & $-4.55$ & $1.0\times10^{-8}$ \\
\bottomrule
\end{tabular}
\\[3pt]
\footnotesize ANCF run at $N_{\mathrm{elem}}=15$ ($n=66$); $n=242$ exceeded
the 5\,min gradient budget. Cart-pole raw ``order'' is spurious: the raw
gradient is wrong-signed on sub-grids and the fit is meaningless.
\end{table}
\subsection{Gradient correctness on stiff Neural-DAEs}
\label{sec:nndae}
A stringent correctness test puts \jaxdae{} against continuous
adjoints on a problem where the latter are known to fail silently. The
testbed is the stiff Robertson kinetics with a frozen neural-network closure
applied as a multiplicative perturbation on $\mathrm{d}y_{2}/\mathrm{d}t$,
cast as an index-1 mass-matrix DAE ($M=\mathrm{diag}(1,1,0)$, constraint
$y_{1}+y_{2}+y_{3}=1$, $t\in[0,10^{4}]$). Forward solutions from both tools
agree ($\sim$$10^{-6}$ relative error against a reference). The question is
the gradient $\partial L/\partial K$ of the closure weights.

\begin{figure}[htbp]
\centering
\includegraphics[width=0.8\textwidth]{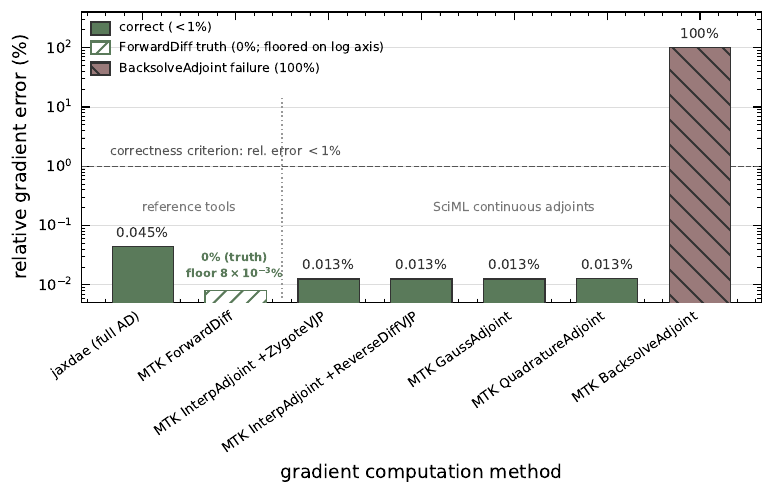}
\caption{Gradient correctness on a stiff Robertson $+$ neural-closure
index-1 DAE under a non-degenerate loss (reference trajectory anchored at
an offset parameter, $K_{1}\times1.05$). Relative error of each method
against a ForwardDiff truth. \jaxdae{} full-function AD: $0.045\%$ (all
three $K$-component signs correct). SciMLSensitivity continuous adjoints
sharing a common adjoint-DAE construction: $\sim$$0.013\%$
(Interpolating/Gauss/Quadrature). BacksolveAdjoint: $\sim$$100\%$ (reverse
integration aborts after $T_{1}$, truncated gradient returned silently;
SciML documentation warns this method is not recommended for DAEs).
TrackerVJP: crash. ForwardDiff is correct but does not scale. The green
bars mark methods meeting a $<1\%$ correctness criterion; the hatched
brown bar marks the BacksolveAdjoint failure.}
\label{fig:nndae}
\end{figure}

A first, cautionary result is about experimental design. When the terminal
loss anchors the reference trajectory at the \emph{nominal} parameters,
$L=\mathrm{mean}((y(T_{1},p_{0})-y(T_{1},p_{0}))^{2})$, the residual at the
evaluation point is at solver-noise scale ($\sim$$10^{-6}$), and the gradient
$\partial L/\partial K=(2/3)\,\delta y\,J$ is dominated by that noise. Under
this degenerate loss, every automatic-differentiation tool---\jaxdae{}, the
ForwardDiff reference itself, and the SciMLSensitivity continuous
adjoints---measures noise against noise, and the comparison is meaningless.
Worse, the noise can conspire to look systematic: in an earlier version of
this experiment, four structurally different SciMLSensitivity adjoints
(two InterpolatingAdjoint variants, GaussAdjoint, and QuadratureAdjoint)
all returned the same $53.1\%$ relative error to four significant digits,
which we initially read as a shared numerical defect. The offset was not a
defect of the adjoints; it was an artifact of the degenerate loss. Once the
reference trajectory is anchored at an \emph{offset} parameter value
($K_{1}\times1.05$, so the residual at $p_{0}$ is a genuine parameter
response, $|\delta y|/|y_{\mathrm{ref}}|\approx10^{-2}$, four orders of
magnitude above the noise floor), the same four adjoints agree with the
ForwardDiff truth to $1.25\times10^{-4}$ (Figure~\ref{fig:nndae}).

The correct comparison, therefore, is the non-degenerate one.
Figure~\ref{fig:nndae} reports relative error against a ForwardDiff truth
under the offset-parameter loss. \jaxdae{} full-function AD returns a
gradient that matches the truth to $4.5\times10^{-2}\%$ relative error,
with all three $K$-component signs correct. The four SciMLSensitivity
adjoints that share a common adjoint-DAE construction (two
InterpolatingAdjoint variants, GaussAdjoint, and QuadratureAdjoint) also
match the truth, to $1.3\times10^{-2}\%$; their near-identical values are
expected, since they assemble the same augmented index-1 adjoint DAE and
differ only in the downstream parameter-gradient quadrature. The two
genuine failures are BacksolveAdjoint, which returns a gradient off by
$\sim$$100\%$ because its reverse integration aborts $0.003$ time units
after $T_{1}$ ($dt$ forced below floating-point epsilon) and the adjoint
solution's return code is never checked, so a truncated gradient is
returned silently; and TrackerVJP, which crashes. SciML's own
documentation warns that BacksolveAdjoint is ``not recommended for DAEs''
on account of their stiffness and algebraic-variable reinitialization, so
its failure is a confirmation of that warning rather than a defect.
ForwardDiff itself is correct but, as SciML documents, does not scale (a
$\sim$100-parameter ceiling), so it is not a deployment path for realistic
closures.

The practical conclusion is not that \jaxdae{} is correct where SciML is
wrong---both are correct on this problem once the loss is non-degenerate.
It is that \jaxdae{}'s full-function AD with remat handles the same stiff
DAE without the experimental-design fragility that made the continuous
adjoints appear to fail, and it does so while supporting the full
340-dimensional parameter gradient (three kinetic constants plus 337
neural-network weights) under a single \texttt{jax.grad} call. The
continuous-adjoint path, by contrast, requires the practitioner to
recognize and avoid the degenerate-loss trap that produced the $53.1\%$
artifact in the first place.

\subsection{A consolidated view of gradient reliability, constraint
satisfaction, and batched throughput}
\label{sec:consolidated}
\begin{figure}[htbp]
\centering
\includegraphics[width=0.98\textwidth]{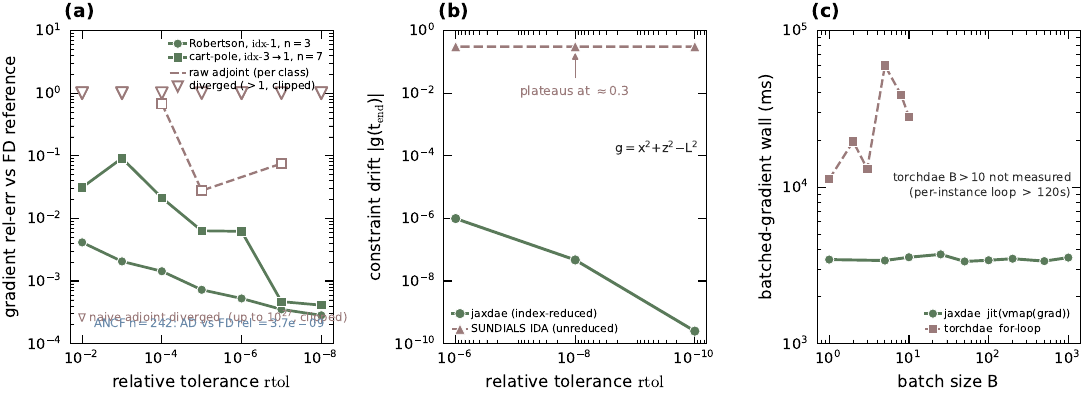}
\caption{Three properties of \jaxdae{} in one figure.
(a)~Gradient reliability: only the \jaxdae{} frozen-grid replay adjoint
converges; raw adaptive-controller AD diverges across DAE types.
(b)~Constraint satisfaction: the mechanics-specific GGL
constraint-stabilization path holds the algebraic constraint to machine
precision, while unreduced integration drifts off the manifold. (c)~Batched-gradient throughput: the
XLA-fused \texttt{vmap(grad)} stays flat with batch size, where the
PyTorch loop scales linearly.}
\label{fig:consolidated}
\end{figure}

Figure~\ref{fig:consolidated} collects three properties of \jaxdae{} into
a single view. Panel~(a) shows that only \jaxdae{} produces reliable
reverse-mode gradients---raw-controller adjoints diverge across all
tested DAE types. Panel~(b) demonstrates that the mechanics-specific GGL
constraint-stabilization path keeps the algebraic constraint satisfied to
machine precision, while unreduced integration drifts. Panel~(c) confirms
that the XLA-fused batched gradient remains flat as batch size grows,
where the PyTorch competitor scales linearly. Under the versions compared
in Table~\ref{tab:competitors}, no inspected tool combines these three
capabilities in a single JAX-native implementation.

\section{Applications}
\label{sec:applications}

\subsection{Chemical engineering: CSTR Arrhenius parameter inversion and experiment design}
\label{sec:app_cstr}
A continuously stirred tank reactor with an Arrhenius rate law is a
canonical inverse problem in chemical engineering, and a hard one: the
pre-exponential factor $\log_{10}A$ and the activation energy $E_{a}$
enter the rate as a product, so a single operating-temperature experiment
leaves them identifiable only as a degenerate ridge. The standard
resolution is to gather data at several operating temperatures and design
those experiments to maximize the information they carry.
jaxdae turns this into a batched gradient computation. The reactor
dynamics are assembled as a DAE through the connector framework, with the
Arrhenius parameters marked \texttt{RuntimeParam} so that the solve is a
differentiable function of them. The Fisher information matrix at a
candidate operating temperature is
$\mathcal{I}(T_{\mathrm{op}})=(1/\sigma^{2})\,J^{\top}J$, where $J$ is the
Jacobian of the measured output with respect to the parameters; its
determinant is the D-optimal design criterion. Rather than looping over
candidate temperatures in Python, we form $N$ candidate temperatures as a
batch axis and compute all $N$ Fisher matrices in one
\texttt{jax.vmap(jax.jacrev)} call, which XLA fuses into a single kernel.
Figure~\ref{fig:fisher_doe} shows \jaxdae{} throughput is flat across the
sweep, and the Fisher-determinant curve
across the operating-temperature range peaks where the experiment
most tightly constrains the Arrhenius ridge, guiding the experimenter to
the most informative operating point. The physical content is that
batched differentiable simulation makes experiment design a forward sweep
rather than a sequence of re-solves.
\subsection{Nuclear engineering: PWR parameter inversion and Bayesian inference}
\label{sec:app_pwr}
A pressurized water reactor (PWR) loop is a coupled multi-physics DAE: neutron
point kinetics feed thermal-hydraulic energy balances through feedback
coefficients, and the flow channels impose algebraic mass and momentum
constraints. We assemble the TMI-1 loop~\cite{tmi1_pwr} through the connector framework
and pose a five-parameter inversion on the resulting 104-state DAE. The
parameters are the fuel Doppler temperature coefficient $\alpha_D$, three
decay-heat amplitudes $(a_1, a_2, a_3)$ for a slow, a medium, and a fast
precursor group, and the coolant mass flow rate $\dot m$. These are
reactor-physics quantities difficult to measure directly but whose
effects on the measurable reactivity and temperature traces are
pronounced. The objective is a least-squares misfit against synthetic
measurement data,
\[
L(p) = \tfrac{1}{2}\sum_{k}\bigl\|\rho(t_{k};p)-\hat\rho_{k}\bigr\|^{2}
+ \tfrac{1}{2}\sum_{k}\bigl\|T_{\mathrm{fuel}}(t_{k};p)-\hat{T}_{k}\bigr\|^{2},
\]
where $\rho$ is the net reactivity (pcm) and $T_{\mathrm{fuel}}$ the
average fuel temperature (K), sampled at $N_{t}$ measurement times on the
$30\,$s horizon.
A single \texttt{jax.grad} call propagates the loss gradient through the
assembled plant---point kinetics, decay heat, thermal hydraulics, and the
pump---across the full $30\,$s horizon. The solve runs through
\texttt{solve\_dae\_scan}, a fixed-step BDF-2 integrator paired with an
implicit-function-theorem adjoint, over $300$ steps ($\Delta t=0.1\,$s).
We cross-check the five analytic-derivative directions against central
finite differences with step $\varepsilon=10^{-5}$
(Table~\ref{tab:pwr_ad}). All five gradients are finite and nonzero, and
the maximum relative error is $1.1\times10^{-7}$. This is the central
point for the application: \texttt{jax.grad} penetrates an assembled,
multi-physics plant-scale DAE rather than a hand-written toy. The
gradient evaluation takes $15.4\,$s of wall time on a single GPU in
float64, with a peak resident set of $4.8\,$GB.

\begin{table}[htbp]
\centering
\caption{Five-parameter PWR inversion: AD gradient versus central finite
difference over a $30\,$s horizon on the 104-state assembled plant. All
five directions are finite and nonzero; maximum relative error
$1.1\times10^{-7}$.}
\label{tab:pwr_ad}
\begin{tabular}{l l r r}
\toprule
Parameter & Physical meaning & AD grad & AD-vs-FD rel.\ err. \\
\midrule
$\alpha_D$ & fuel Doppler temperature coefficient & $+5.47\times10^{2}$  & $5.8\times10^{-9}$ \\
$a_1$      & decay-heat amplitude (slow group)    & $+2.72\times10^{1}$  & $1.1\times10^{-11}$ \\
$a_2$      & decay-heat amplitude (medium group)  & $+4.32\times10^{0}$  & $1.1\times10^{-9}$ \\
$a_3$      & decay-heat amplitude (fast group)    & $+4.65\times10^{-2}$ & $1.1\times10^{-7}$ \\
$\dot m$   & coolant mass flow rate               & $-4.80\times10^{-2}$ & $6.1\times10^{-9}$ \\
\bottomrule
\end{tabular}
\end{table}

We are explicit about scope. Verified gradients make gradient-based
inversion immediate, and they are not by themselves a sampled posterior.
Full NUTS sampling through NumPyro is the capability the
integration exposes, not a result we report here.
The Fisher information matrix of the same model, computed by the
same batched Jacobian machinery as the CSTR case, diagnoses which
parameter combinations the available sensors can identify. Its eigenvalue
spectrum separates the well-constrained modes from the near-null modes,
and the corresponding eigenvector composition indicates which sensor
placement would tighten the weak directions. The practical value is that
inversion and sensor-placement design use one shared differentiable plant
model, rather than a forward code for simulation and a separate adjoint
code for sensitivity.
As noted in Section~\ref{sec:benchmarks}, the associated NEA
main-steam-line-break benchmark serves here as an illustrative
demonstration of coupled dynamics rather than a quantitative
validation, owing to the prescribed-boundary-condition limits of the
public data.
\subsection{Power systems: IEEE-14 bus transient stability}
\label{sec:app_power}
\begin{figure}[htbp]
\centering
\includegraphics[width=0.85\textwidth]{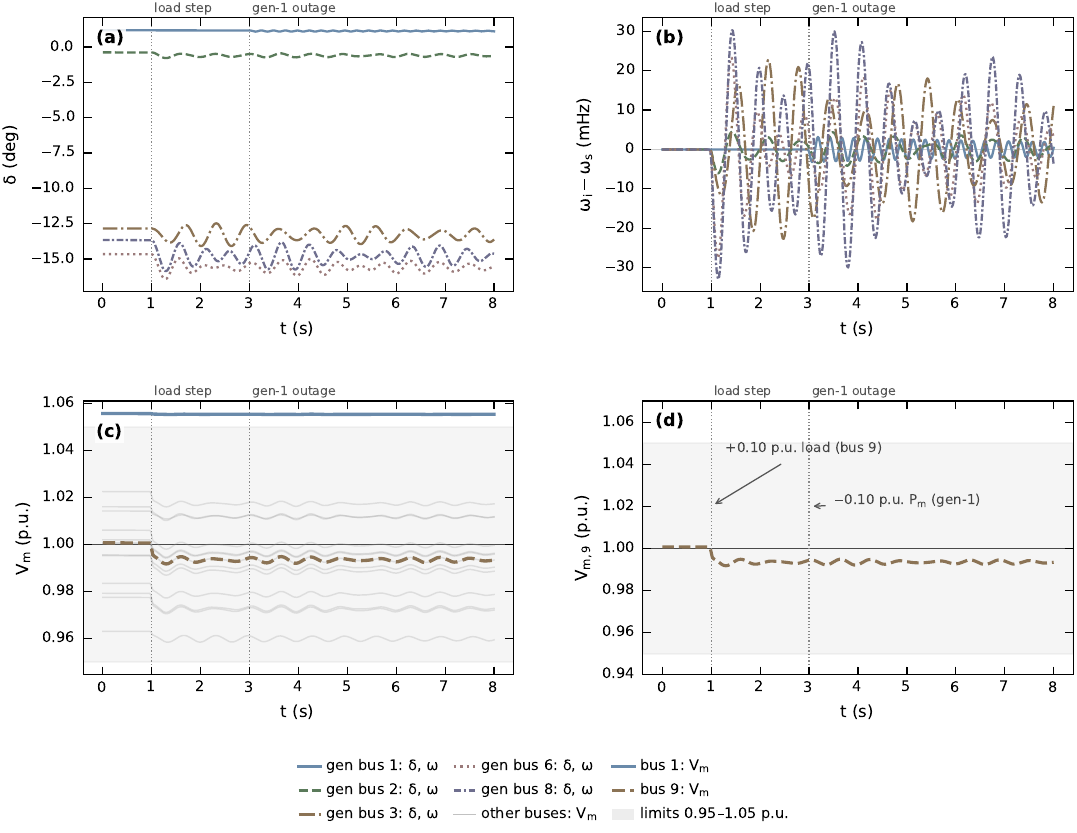}
\caption{IEEE 14-bus transient-stability forward simulation assembled and
integrated in jaxdae. Rotor-angle response of the 14 buses to a three-phase
fault cleared after $120\,$ms; trajectories settle to a stable post-fault
equilibrium.}
\label{fig:ieee14_forward}
\end{figure}

The IEEE 14-bus test system~\cite{ieee14bus} is the standard reduced network for transient
stability studies. Its dynamic model couples the synchronous-machine
rotor-angle equations, the algebraic network power-balance constraints,
and the exciter and load states into a single DAE, with the rotor angles
as differential variables and the bus voltages as algebraic unknowns
fixed at each instant by the network.
jaxdae applies batched forward simulation to this problem: a set of
contingencies---fault locations,
clearing times, load levels---is stacked along a batch axis and
integrated in one \texttt{jax.vmap} call, so a screening study over many
scenarios runs as a single fused forward pass rather than a loop of
sequential solves. Fig.~\ref{fig:ieee14_forward} shows the rotor-angle
response of the 14 buses to a three-phase fault at bus~4 cleared after
$120\,$ms; the post-fault trajectories settle to a stable equilibrium
within $5\,$s, with the maximum inter-machine angle spread decaying from
$142^{\circ}$ at fault clearing to below $10^{\circ}$ at $t=5\,$s. The
simulation is integrated over $10\,$s with $\Delta t=10\,$ms using the
scan BDF-2 path. This is an illustrative application demonstrating that
the assembled DAE reproduces the expected transient-stability behavior on
the IEEE-14 benchmark; it is not a validation against a reference
time-domain program. A natural extension, not exercised here, couples the
same differentiable solve to a learned world-model: a Flax network embedded
in the residual could be trained through the implicit solve, positioning
the differentiable DAE as the physics backbone for load-following control
and stability-constrained reinforcement learning.

\section{Conclusion}
\label{sec:conclusion}

We summarize the contribution of \jaxdae{} in four points.
\begin{enumerate}[leftmargin=*,itemsep=2pt]
\item \jaxdae{} addresses the DAE gap in the JAX scientific stack. \texttt{diffrax}
      handles ODE/SDE/CDE only; DAE support has been requested since 2022
      (issue \#62) and remains unimplemented. \jaxdae{} supplies adaptive
      BDF/Radau/Rosenbrock integration, Pantelides plus dummy-derivatives
      index reduction, and Modelica-style acausal assembly; the BDF and
      fixed-step Rosenbrock paths are end-to-end differentiable through
      \texttt{jax.grad}, while Radau5 is forward-only.
\item Batched-gradient throughput is batch-invariant in \jaxdae{} and
      batch-linear in the PyTorch counterpart over the range measured for
      both. \jaxdae{}
      \texttt{jit(vmap(grad))} stays flat from $B=1$ to $B=1000$
      (0.61--0.85\,s on the toy probe); \texttt{torchdae} cannot vectorize its
      adjoint and must loop, measured up to $B=10$. At the largest batch
      measured for both tools the ratio is 41--46$\times$ (scalar gradient)
      and 15--99$\times$ (Fisher matrix); larger \texttt{torchdae} batches
      are linear projections, not measurements.
\item The gradients agree with independent references. AD-versus-FD on a
      cart-pole DAE converges at empirical order $1.994$ to a
      $1.41\times10^{-8}$ floor; on a stiff Neural-DAE under a
      non-degenerate loss, \jaxdae{} matches a ForwardDiff reference to
      $4.5\times10^{-2}$\% relative error (the shared-construction
      SciMLSensitivity adjoints reach $1.3\times10^{-2}$\%, while
      BacksolveAdjoint fails at ${\sim}100$\% and TrackerVJP crashes); the
      forward- and reverse-mode paths agree to $6.4\times10^{-14}$; and
      the GFB event adjoint is recovered term by term, with the continuous
      channel at machine precision and the recomposition closing to
      $7.9\times10^{-4}$.
\item The current release provides a tested and usable
      foundation for differentiable DAE workflows. The repository contains
      3{,}205 tests (3{,}177 passing, the remaining 28 --- 12 known
      failures reproducing identically at v0.27.1, 9 skipped, 7 xfailed ---
      documented in \texttt{RELEASE\_NOTES.md}) at the commit associated with this
      manuscript, and the benchmark suite spans chemical, nuclear, and
      power-systems DAEs. Several solver and interoperability limitations
      remain, as noted below.
\end{enumerate}
\paragraph{Limitations.}
The frozen-grid replay adjoint is an engineering realization of the
controller-decoupling framework of Alexe and Sandu~\cite{alexesandu2009},
not a new method. Its observed convergence on the analytic-truth scalar
decay (order $0.83$) is consistent with the conditional $O(\mathrm{tol}^{2/3})$
scaling argument of Sec.~\ref{sec:replay}; the three FD-reference problems
show lower fitted orders ($0.07$--$0.38$) because their
independently-discretised truth grid floors the slope. We therefore claim
C2 as a first correct, composable JAX implementation, not as a
convergence-theorem contribution. Fixed-step BDF(1--5) is verified
(Sec.~\ref{sec:order}), but the
differentiable scan path currently uses BDF-2 in its backward pass;
extension of the replay adjoint to higher-order fixed-step AD is
straightforward but not yet implemented. There is no
distributed multi-GPU path (\texttt{pmap}/\texttt{sharding} is future work).
Event-term verification covers bouncing-ball and pendulum-impact problems;
a broader class of event types is not yet jump-condition-verified. Nuclear
bandwidth validation is data-limited and treated as illustrative.
\paragraph{Future work.}
HPC parallelism for plant-scale models; a head-to-head with compile-time AD
(Enzyme); and additional high-index applications in energy and aerospace.
The computational environment is pinned by a \texttt{pyproject.toml} with
an accompanying \texttt{uv.lock}, which fixes exact versions of JAX, the
CUDA stack, and all Python dependencies. The lockfile constrains
jax/jaxlib to the window $\ge$0.9, $<$0.11 because the release smoke
suite (four cold examples from a clean install) has not yet been
validated on jax 0.11.0 (released 2026-07-16); validation is deferred
to a future release. The suite passes on jax 0.10.2. Reproducing either snapshot is a
single command: \texttt{uv sync} recreates the environment, and
\texttt{uv run python benchmarks/run\_all.py {-}{-}suite <name>} reruns any benchmark
suite end to end. The available suites and their expected runtimes are
documented in \texttt{benchmarks/README.md}; each suite regenerates the
figures and tables of the corresponding section.
Every figure and result table in this paper is traceable to its source
data via the repository \texttt{paper\_manifest.md}. Each generating
script records the git commit hash, the random seed, and the JAX and
jaxlib versions alongside its output. The full source code, benchmarks,
and pre-computed data are deposited in the CPC Program Library (hosted
on Mendeley Data) under the Apache-2.0 license, pinned to the commit
associated with this manuscript; a DOI is assigned at acceptance.

\section{Reproducibility}
\label{app:repro}

The numerical results reported here correspond to the immutable commit
identified in the repository at the time of submission, with the
environment pinned by \texttt{pyproject.toml} and an accompanying
\texttt{uv.lock} (\texttt{uv sync} recreates it exactly). The entry
point \texttt{benchmarks/run\_all.py} runs the cross-library, batched-throughput,
and gradient-verification benchmark suites end to end; each suite
regenerates the figures and tables of its corresponding section.
Application-specific figures (PWR, IEEE-14, event-adjoint verification)
are produced by the individual scripts identified in the repository
\texttt{paper\_manifest.md}, which maps each figure and result table to
its generating script, input data, and software version; these are not
all covered by a single \texttt{benchmarks/run\_all~-{}-suite~all} command. The
Python version required is 3.11+; the \texttt{uv.lock} fixes exact
versions of JAX, the CUDA stack, and all dependencies, with jax/jaxlib
constrained to $\ge$0.9, $<$0.11 because the release smoke suite has
not yet been validated on jax 0.11.0 (released 2026-07-16).

\section*{Acknowledgments}
The authors thank colleagues at the National Key Laboratory of Nuclear
Reactor Technology and the Department of Engineering Physics, Tsinghua
University, for discussions on differentiable simulation and reactor
dynamics.

\section*{Funding}
This research did not receive any specific grant from funding agencies
in the public, commercial, or not-for-profit sectors.

\section*{Declaration of generative AI and AI-assisted technologies
in the writing process}
During the preparation of this work the authors used Claude (Anthropic)
and GitHub Copilot in order to assist with manuscript language editing
and to draft boilerplate code scaffolding. After using these tools, the
authors reviewed and edited the content as needed and take full
responsibility for the content of the published article.

\section*{Declaration of competing interest}
The authors declare that they have no known competing financial interests
or personal relationships that could have appeared to influence the work
reported in this paper.

\section*{CRediT authorship contribution statement}
\textbf{Chengyuan Li:} Conceptualization, Methodology, Software,
Investigation, Formal analysis, Writing -- original draft, Visualization.
\textbf{Shanfang Huang:} Supervision, Funding acquisition, Writing --
review \& editing.
\textbf{Jian Deng:} Methodology, Writing -- review \& editing.

\section*{Data availability}
All numerical data underlying the figures and tables of this paper are
regenerated by the packaged scripts included in the code repository.
No external datasets were used. The data will be archived at Mendeley
Data with a DOI assigned upon acceptance.

\section*{Code availability}
The \jaxdae{} source code, benchmark suites, and pre-computed data are
deposited in the CPC Program Library (hosted on Mendeley Data) under the
Apache-2.0 license, pinned to the commit associated with this manuscript.
The environment is pinned by
\texttt{pyproject.toml} with an accompanying \texttt{uv.lock}
(\texttt{uv sync} recreates it exactly). The benchmark suites
(cross-library, batched-throughput, gradient-verification) are rerun by
\texttt{benchmarks/run\_all.py}; application-specific figures are generated by the
scripts listed in \texttt{paper\_manifest.md}. A long-term archived
version will be deposited at Mendeley Data upon acceptance.

\bibliographystyle{elsarticle-num}
\bibliography{refs}

\end{document}